\documentclass[11pt]{article}
\pdfoutput=1

\usepackage{jheppub}
\usepackage{url,comment}
\usepackage{times}
\usepackage{latexsym}
\usepackage{graphicx, graphics, hyperref, amsmath, amssymb, slashed, xcolor, bbm,bm,amsthm, array}
 \usepackage{subfigure}
 \usepackage{listings} 
 \usepackage{trimclip}

\usepackage{rotating}
\usepackage{afterpage}
\usepackage{multirow}

\usepackage{lineno}
\newcommand{\nc}{\newcommand}
\nc{\ttbar}{t\bar t}
\nc{\bbbar}{b\bar b}
\nc{\ccbar}{c\bar c}

\graphicspath{{plots/}}

\title{The \mbox{MATHUSLA} Test Stand}

\author[a]{Maf Alidra,}
\author[b]{Cristiano Alpigiani,}
\author[a]{Austin Ball,}
\author[c,d]{Paolo Camarri,}
\author[c]{Roberto Cardarelli,}
\author[e]{John Paul Chou,}
\author[f]{David Curtin,}
\author[g]{Erez Etzion,}
\author[e]{Ali Garabaglu,}
\author[e]{Brandon Gomes,}
\author[a]{Roberto Guida,}
\author[b]{W. Kuykendall,}
\author[b]{Audrey Kvam,}
\author[h]{Dragoslav Lazic,}
\author[b]{H. J. Lubatti,}
\author[i, j]{Giovanni Marsella,}
\author[g]{Gilad Mizrachi,}
\author[k]{Antonio Policicchio,}
\author[b]{Mason Proffitt,}
\author[b]{Joe~Rothberg,}
\author[c,d]{Rinaldo Santonico,}
\author[g]{Yiftah Silver,}
\author[e]{Steffie Ann Thayil,}
\author[l,*]{Emma~Torro-Pastor,}
\author[b]{Gordon Watts,}
\author[m]{Charles Young}

\affiliation[a]{CERN, Switzerland}
\affiliation[b]{University of Washington, Seattle, USA}
\affiliation[c]{Istituto Nazionale di Fisica Nucleare, Sezione di Roma Tor Vergata, Roma, Italy}
\affiliation[d]{Universit\`{a} degli Studi di Roma Tor Vergata, Roma, Italy}
\affiliation[e]{Rutgers University, USA}
\affiliation[f]{University of Toronto, Canada}
\affiliation[g]{Tel Aviv University, Israel}
\affiliation[h]{Boston University, USA}
\affiliation[i]{Universit\`{a} del Salento, Lecce, Italy}
\affiliation[j]{Istituto Nazionale di Fisica Nucleare, Sezione di Lecce, Lecce, Italy}
\affiliation[k]{Universit\`{a} degli Studi di Roma La Sapienza, Roma, Italy}
\affiliation[l]{Instituto de F\'isica Corpuscular (CSIC-UV), Valencia, Spain}
\affiliation[m]{SLAC National Accelerator Laboratory, USA}

\affiliation[*]{Corresponding author. Email address: Emma.Torro.Pastor@cern.ch}

\abstract{

The rate of muons from LHC $pp$ collisions reaching the surface above the ATLAS interaction point is measured as a function of the ATLAS luminosity and compared with expected rates from decays of $W$ and $Z$ bosons and $b$- and $c$-quark jets.  
In addition, data collected during periods without beams circulating in the LHC provide a measurement of the background from cosmic ray inelastic backscattering that is compared to simulation predictions.  
Data were recorded during 2018 in a 2.5 $\times$ 2.5 $\times$ 6.5~$\rm{m}^3$ active volume MATHUSLA test stand detector unit consisting of two scintillator planes, one at the top and one at the bottom, which defined the trigger, and six layers of RPCs between them, grouped into three $(x,y)$-measuring layers separated by 1.74 m from each other. Triggers selecting both upward-going tracks and downward-going tracks were used. 

}


\begin{document}

\maketitle

%%%%%%%%%%%%%%%%%
\section{Introduction}
\label{s.introduction}
%%%%%%%%%%%%%%%%%

A small-scale experiment, the \mbox{MATHUSLA} test stand, was constructed and installed on the surface above the interaction point (IP) of the \mbox{ATLAS} detector at Point 1 of the LHC and collected data during 2018.  
The detector was operational both during LHC $pp$ collisions and when the LHC was not colliding protons. 
The goal was to measure the rate of muons from LHC $pp$ collisions reaching the surface, as well as the rate of inelastic backscattering from cosmic rays that could create upward-going tracks, and to determine how well simulation models could reproduce the data. 
This information will be a very useful input for future studies on the background expectations for the proposed \mbox{MATHUSLA} (MAssive Timing Hodoscope for Ultra-Stable neutraL pArticles) detector \cite{Chou:2016lxi,Alpigiani:2018fgd}. 

The test stand used scintillation counters recovered from the Tevatron Run II D\O\, forward muon trigger system \cite{Abazov_2005}.
The scintillators, arranged to cover two planes of a 2.5~$\times$~2.5~m$^2$ area each, were used to form the test stand trigger. 
Spare resistive plate chambers (RPCs) originally built for the ARGO-YBJ experiment \cite{ref:ARGO}, arranged in six layers between the scintillator planes, were used to track charged particles traversing the test stand.

The test stand is described and discussed in Section 2.  Section 3 presents the details of timing calibration, track reconstruction, and detector efficiency estimation, and Section 4 describes simulation of expected events from cosmic rays and LHC $pp$ collisions. The measured results and comparisons to simulation predictions are shown and discussed in Section 5.

\section{\mbox{MATHUSLA} test stand description}
\label{s.teststand}
%%%%%%%%%%%%%%%%%

\begin{figure}[h]
\label{f.mathuslateststand}
\begin{center}
\subfigure[]{
\label{f.mathuslateststand_a}
\includegraphics[height=7cm]{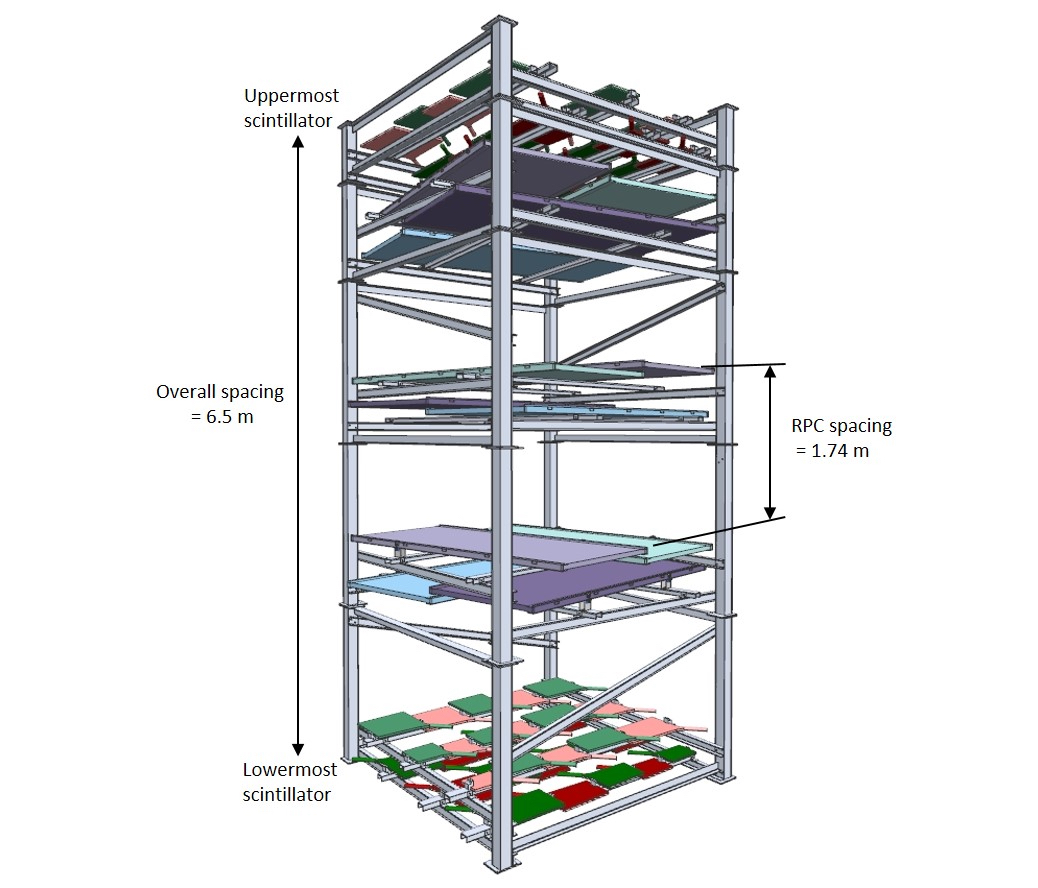}
}
\subfigure[]{
\label{f.mathuslateststand_b}
\includegraphics[height=7cm]{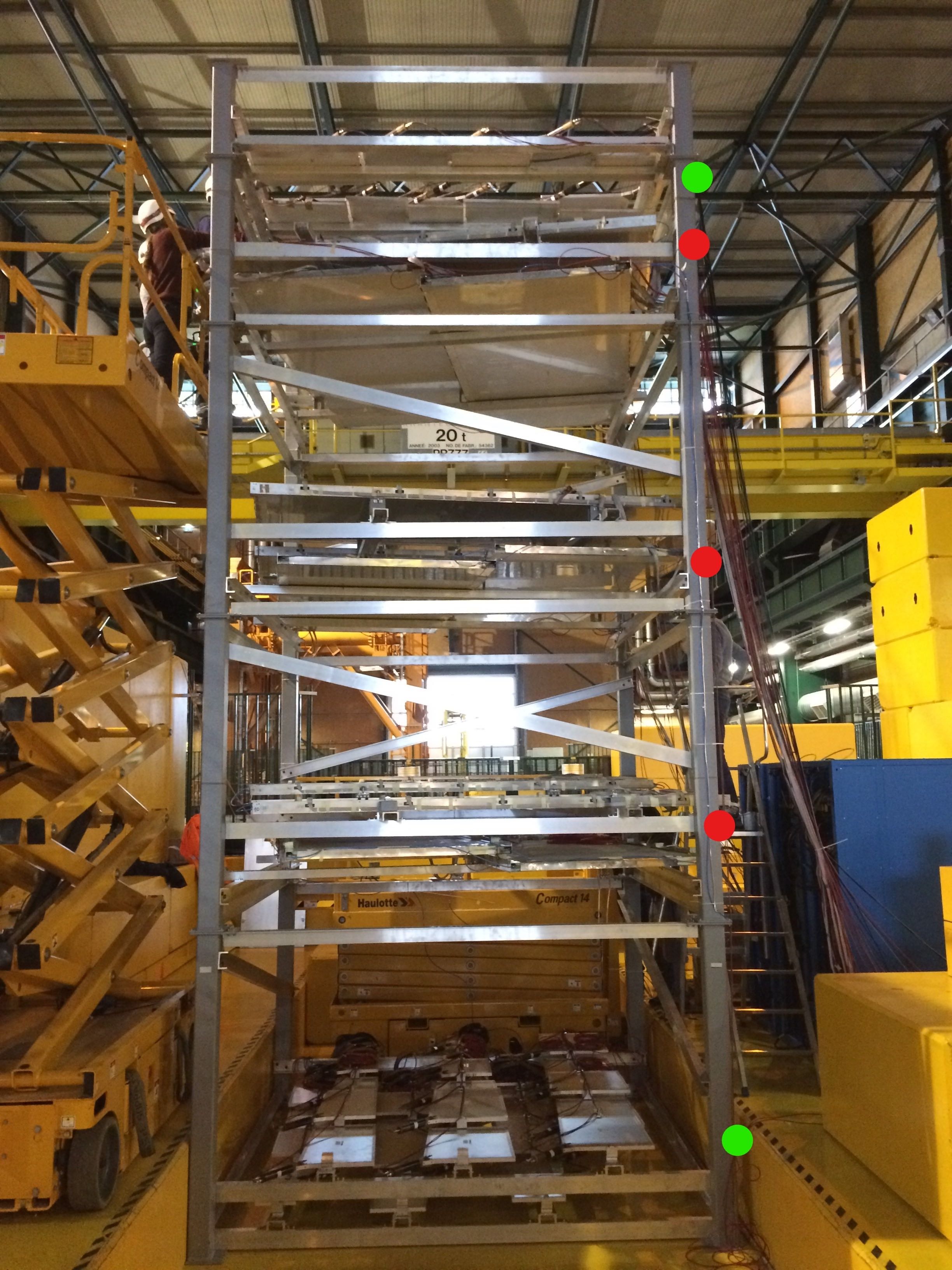}
}
\end{center}
\caption{
(a) 3D model of the \mbox{MATHUSLA} test stand. (b) Photo of the final assembled structure installed above the ATLAS IP. The green dots identify the two scintillator layers used for triggering, while the red dots mark the three RPC double-layers used for tracking.
}

\end{figure}

The \mbox{MATHUSLA} test stand comprised two planes of scintillation counters, one at the top and one at the bottom, with six layers of RPCs between them that were grouped into three double-layers.
Figure~\ref{f.mathuslateststand_a} shows the basic design of the test stand.
The overall structure was 6.8~m tall with a distance of 6.5~m between the lowermost and uppermost scintillators. The three RPC double-layers were located at approximately 2~m, 3.7~m, and 5.5~m above the lowermost scintillators.
The structure had a base of 2.98~$\times$~2.91~m$^2$ with an active area of approximately 2.5~$\times$~2.5~m$^2$. 
Figure~\ref{f.mathuslateststand_b} shows the test stand in the SX1 building at CERN, 80~m above the ATLAS IP.

The scintillator planes were used to form the trigger for the test stand, while the RPC layers were used to measure spatial and time coordinates for tracking. The following subsections contain further information on the test stand subdetectors, trigger, and data acquisition systems.

%%%%%%%%%%%%%%%%%%%%%%%%%%%%
\subsection{Scintillation counters}

\begin{figure}[h]
\vspace{-0.75cm}
\begin{center}
\begin{minipage}[r]{0.45\textwidth}
\includegraphics[width=\textwidth]{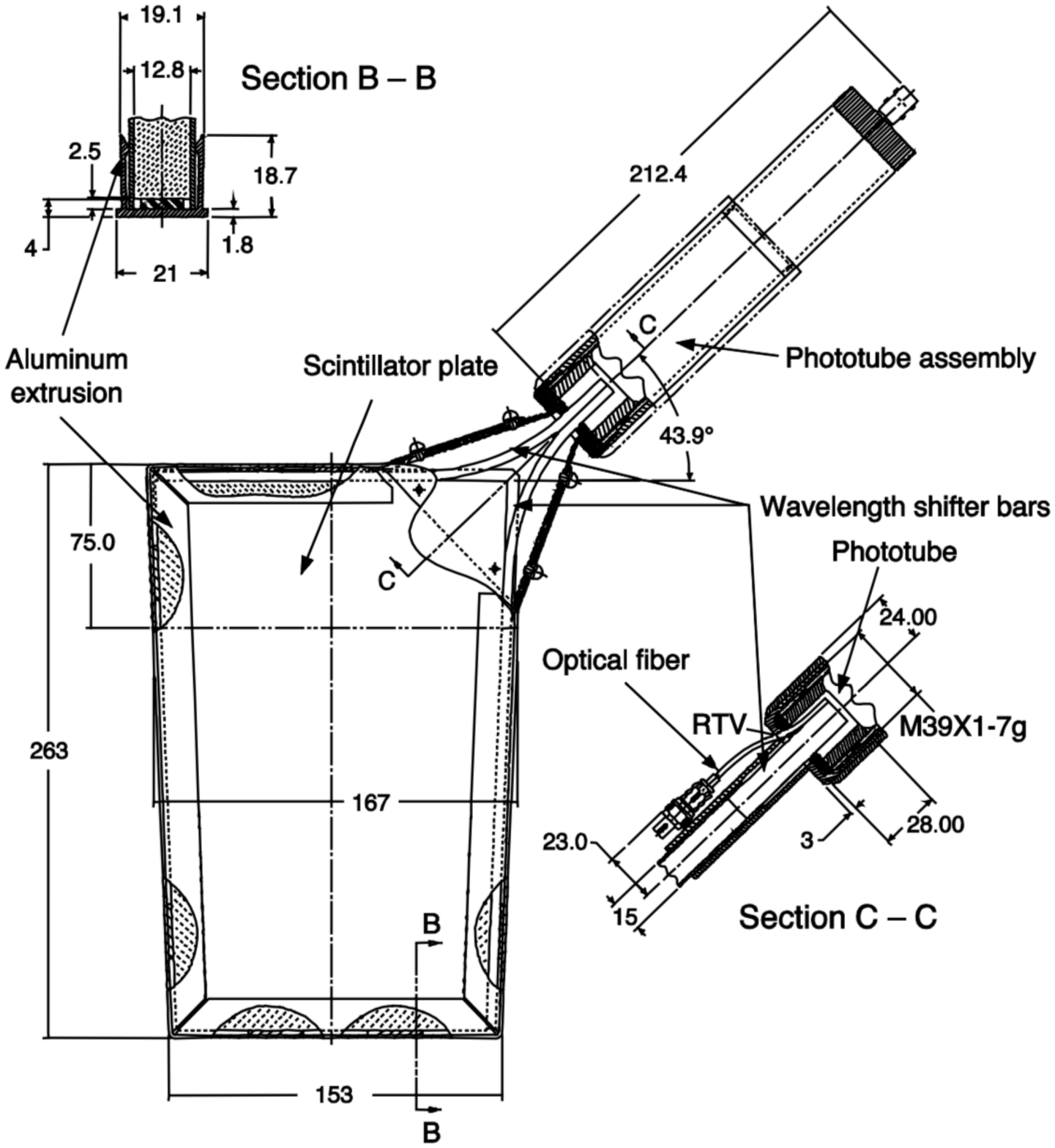}
\end{minipage}
\hspace{0.05\textwidth}
\begin{minipage}[l]{0.45\textwidth}
%\vspace{-0.8 cm}
\includegraphics[width=\textwidth]{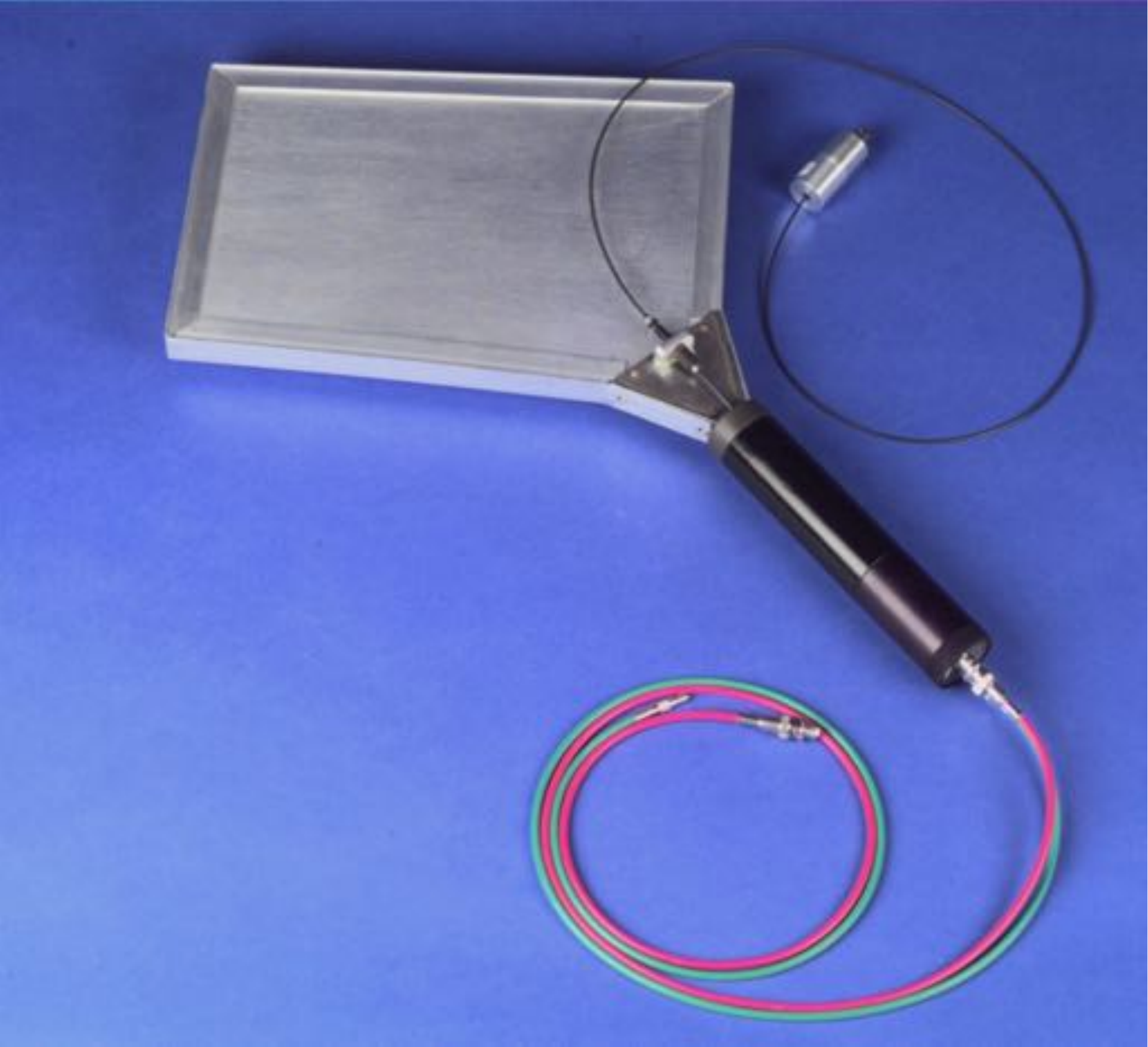}
\end{minipage}
\end{center}
\vspace{-1.25cm}
\caption{
Schematic \cite{Abazov_2005} and photo of a D\O\, forward muon scintillation trigger counter.
}
\label{f.scintillator}
\end{figure}

The scintillators used in the \mbox{MATHUSLA} test stand are spare scintillation counters from the forward muon trigger system of the D\O\, detector at the Tevatron at Fermilab \cite{Abazov_2005}.
The scintillator tiles are made of 12.7~mm-thick BICRON 404A plastic, which has a light emission peak at 420~nm and an attenuation length of 1.7~m. 
Each tile has two wavelength-shifting (WLS) bars with an absorption peak matching the emission peak of the scintillator. 
As shown in Figure \ref{f.scintillator}, the WLS bars are located on two edges of each scintillator and double as light guides. 
The bars are made of SOFZ-105, based on PMMA (polymethylmethacrylate) plastic,  and contain the wavelength-shifting fluorescent dopant Kumarin~30. 
One end of each bar directs the light signal into a 25~mm-diameter MELZ 115M photomultiplier tube (PMT). 
The sensitivity peak of the PMTs matches the 480 nm emission peak of the WLS bars. 
At the end opposite to the PMT there is a mylar tape reflector. 
The scintillator tile and WLS bars are wrapped in TYVEK type 1056D and photographic paper to ensure light-tightness and are encased in an aluminum outer shell.
Each PMT is connected to a high-gain base and is surrounded by a magnetic shielding tube.

\subsubsection{Counter testing}

The scintillator tiles and PMTs were tested with cosmic rays before being installed in the test stand. 
The dark noise of all available PMTs was measured as a function of supply voltage in the range \mbox{1.8--2.2 kV}. Noise hits were required to pass a 30 mV discriminator threshold. 
The PMTs with the least noise were then paired with scintillator tiles and tested at different supply voltages to ensure a charged particle detection efficiency $\geq$ 97\% at the corner of the scintillator farthest from the PMT. 

\subsubsection{Scintillator plane assembly}

\begin{figure}[h]
\begin{center}
\subfigure[]{\label{f.mathuslateststand:a}\includegraphics[height=5.2cm]{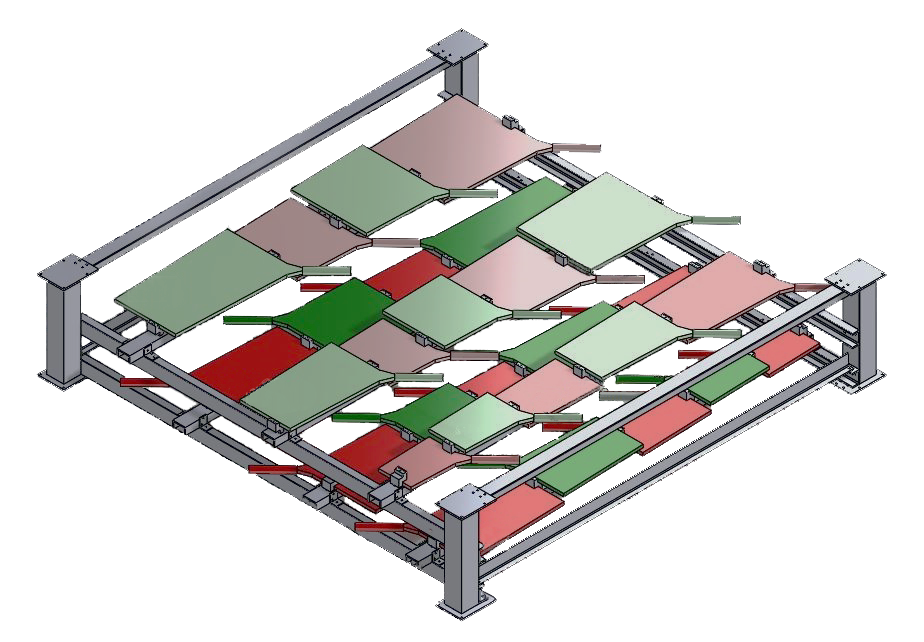}\label{f.scintillatorLayout_a}}
\subfigure[]{\label{f.mathuslateststand:b}\includegraphics[height=5.8cm]{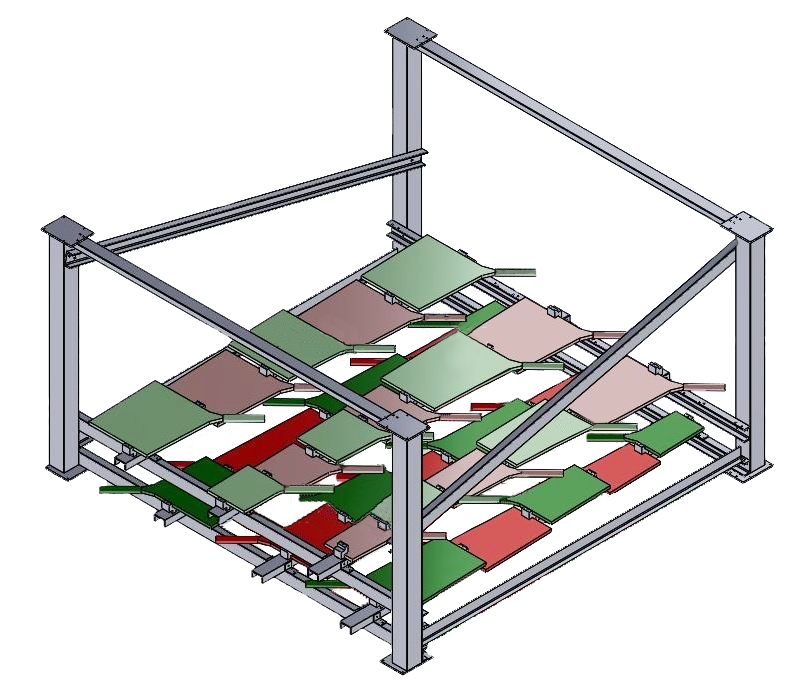}\label{f.scintillatorLayout_b}}
\end{center}
\caption{
Layout of the top (a) and bottom (b) scintillator planes, each of which has a 2.5 $\times$ 2.5~$\rm{m}^2$ active area.
}
\label{f.scintillatorLayout}
\end{figure}

The counters were arranged into two approximately square planes with an area of 2.5 $\times$ 2.5 m$^2$ each. The counters were placed in rows at different heights as shown in Figure \ref{f.scintillatorLayout} to allow for overlap in order to avoid gaps in the area coverage. 
The top and bottom scintillator planes were composed of 28 and 31 counters, respectively. 
The size of the smallest counters is approximately 22~$\times$~37~cm$^2$ while the largest counters measure 63~$\times$~69~cm$^2$.

%%%%%%%%%%%%%%%%%%%%%%%%%%%%
\subsection{Resistive plate chambers (RPCs)}
\label{sec:rpc}

\begin{figure}[h]
\begin{center}
\subfigure[]{\label{f.rpc:a}\includegraphics[height=4.9cm]{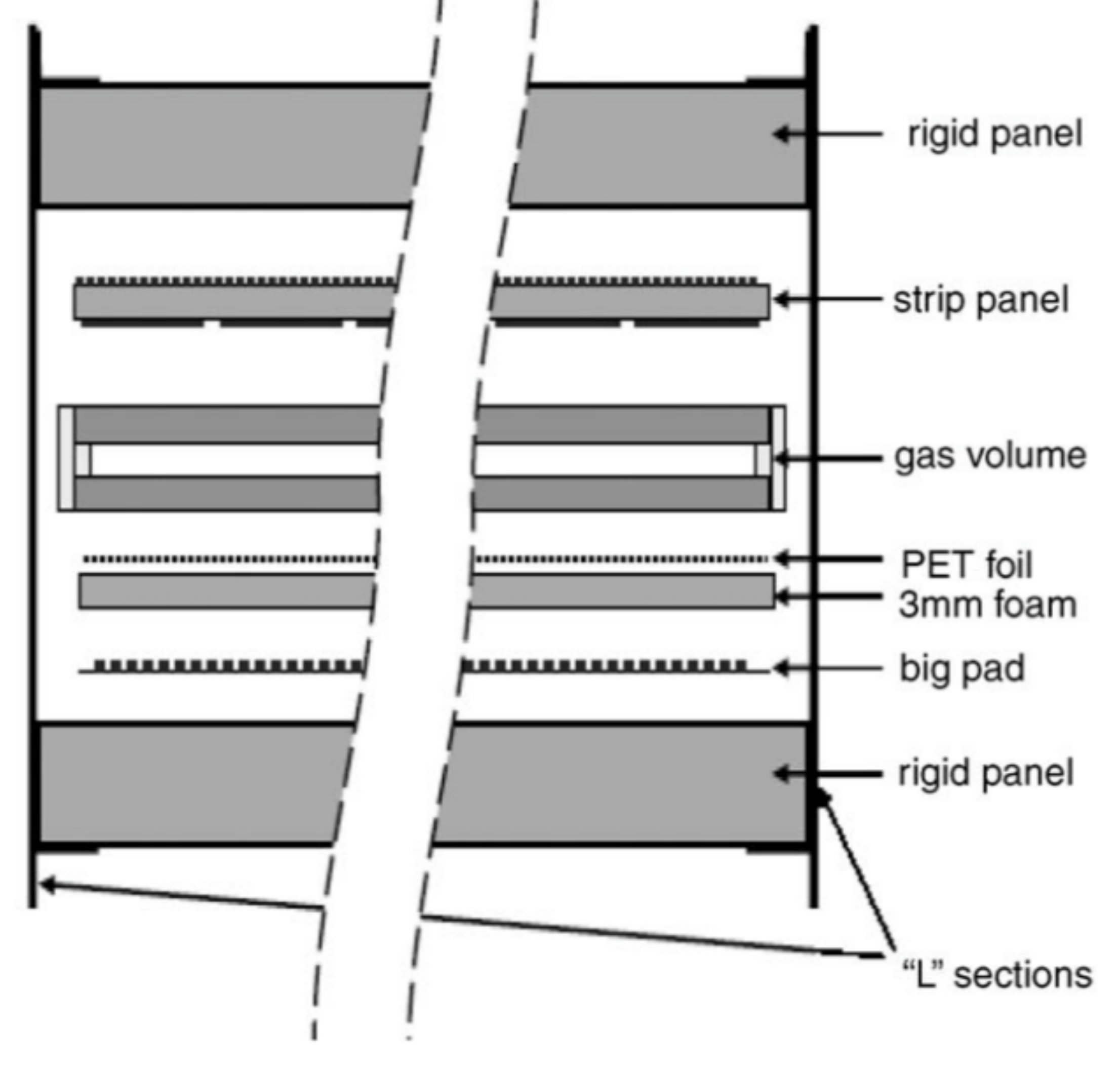}}
\subfigure[]{\label{f.rpc:b}\includegraphics[height=4.9cm]{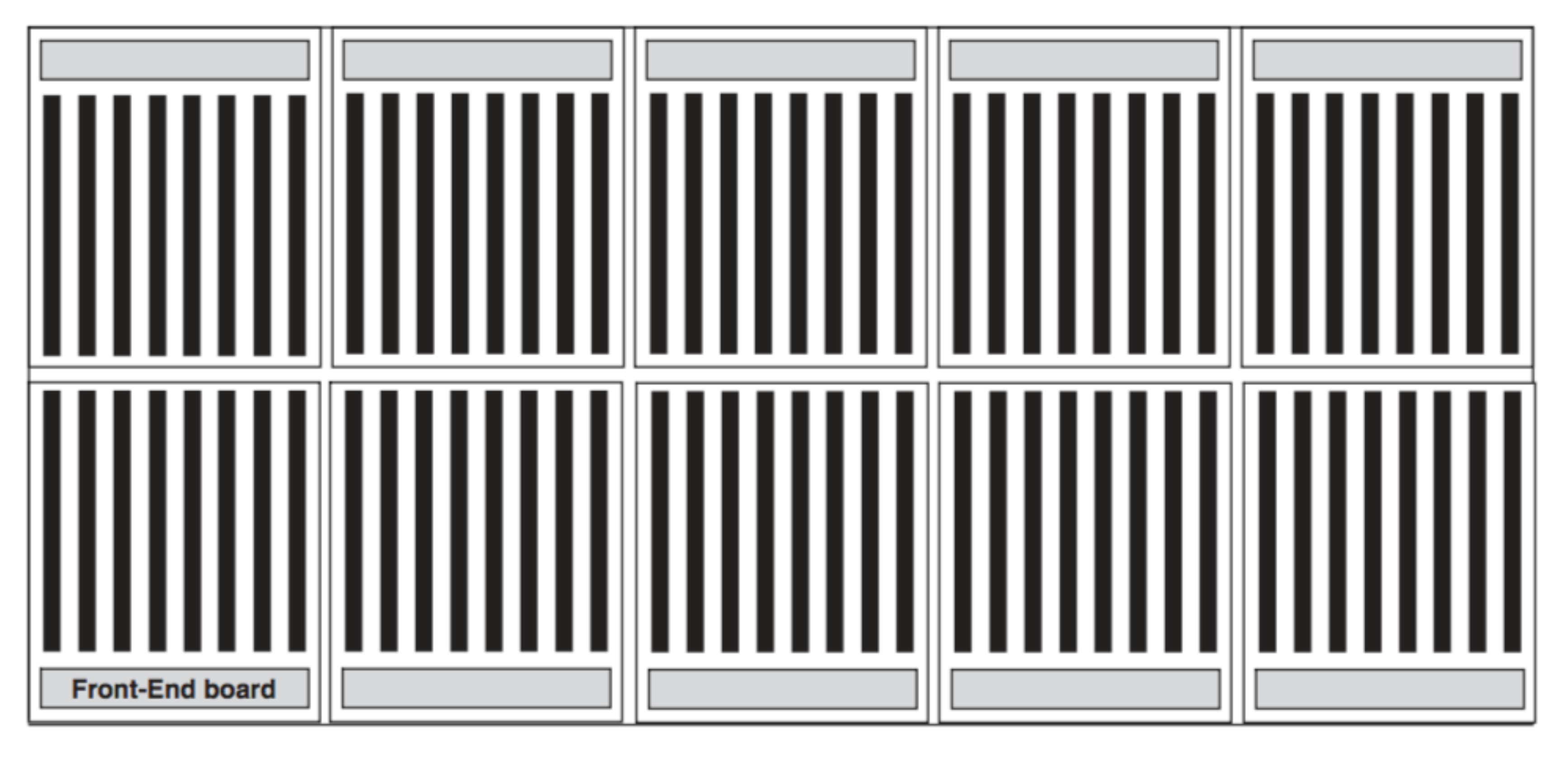}}
\end{center}
\caption{
(a) Schematic of the cross-section of an ARGO-YBJ chamber \cite{ref:ARGORPC}. (b) Strip panel used for readout. The total size of each chamber is 285~$\times$~126~$\times$~4.7~cm$^3$.
}
\label{f.rpc}
\end{figure}

Spare RPCs from the ARGO-YBJ experiment~\cite{ref:ARGO} were used for the MATHUSLA test stand.
Each chamber consists of a 2 mm-thick gas gap with a sensitive area of 2.70~$\times$~1.23~m$^2$ and a readout strip panel, both assembled inside a 47 mm-thick Faraday cage that also serves as a mechanical support for the chamber.
The readout panel can pick up the signals generated inside the gas gap by means of 80 copper strips of 6.76~$\times$~62.35~cm$^2$.
Figure \ref{f.rpc} shows a cross-section view of an ARGO-YBJ chamber and a sketch of the strip panel used for readout. 
The front-end boards are soldered at the end of the strips and embedded in the Faraday cage. 
Eight contiguous strips form a pad with a size of 55.68~$\times$~62.35~cm$^2$.
The pad signal is the logical OR of the eight strips and is used for timing.

For the \mbox{MATHUSLA} test stand, it was required that the RPCs be able to track both LHC muons and cosmic rays. 
The ARGO-YBJ chambers were originally designed for counting cosmic ray shower particles and not for tracking. 
Consequently the strip size was not optimized for spatial resolution.
Additionally, in ARGO-YBJ the chambers operated in streamer mode with a gas mixture of 75\% tetrafluoroethane, 15\% argon, and 10\% isobutane \cite{ref:ARGORPC}.
RPCs operating in streamer mode have been used for tracking in the past \cite{ref:MINI}. 
In the test stand, the RPCs were operated in streamer mode using the standard ATLAS RPC gas mixture (94.7\% tetrafluoroethane, 5\% isobutane, and 0.3\% sulfur hexafluoride \cite{ref:atlasrpc}) with an addition of 15\% of argon.

\subsubsection{RPC high voltage correction}

The high voltage applied to the RPCs was continuously adjusted so that the effective voltage $V_{\mathrm{eff}}$, which determines the gas gain, remained constant as a function of chamber temperature $T$ and  atmospheric pressure $p$. $V_{\mathrm{eff}}$ is given by the following formula \cite{ref:MINI}:
\begin{linenomath*}
\begin{equation*}
V_{\mathrm{eff}} = V_{\mathrm{app}} \frac{T}{T_0} \frac{p_0}{p}
\end{equation*}
\end{linenomath*}
where $V_{\mathrm{app}}$ is the applied voltage and $T_0$ and $p_0$ are reference temperature and pressure values, respectively. 
Therefore, for a given time $t$, the applied voltage $V_{\mathrm{app}}$ that ensures $V_{\mathrm{eff}}$ remains constant and equal to the reference voltage $V_0$ is:
\begin{linenomath*}
\begin{equation*}
V_{\mathrm{app}}(t) = V_0 \frac{T_0}{{p_0}} \frac{p(t)}{T(t - 1 \textrm{ hr})}
\end{equation*}
\end{linenomath*}
The effects on the voltage due to the pressure changes are immediate, but there is a delay for a temperature change occurring outside the chamber to be reflected in the gas temperature inside the RPC. For the ARGO-YBJ chambers, the delay is one hour \cite{Camarri_2013}.  
A BME 280 Bosch sensor attached to an Arduino Uno board was used to measure the temperature and the atmospheric pressure.
This environmental information was used to adjust the voltage every five minutes.

\subsubsection{RPC layer assembly}

\begin{figure}[h]
\begin{center}
\includegraphics[height=6cm]{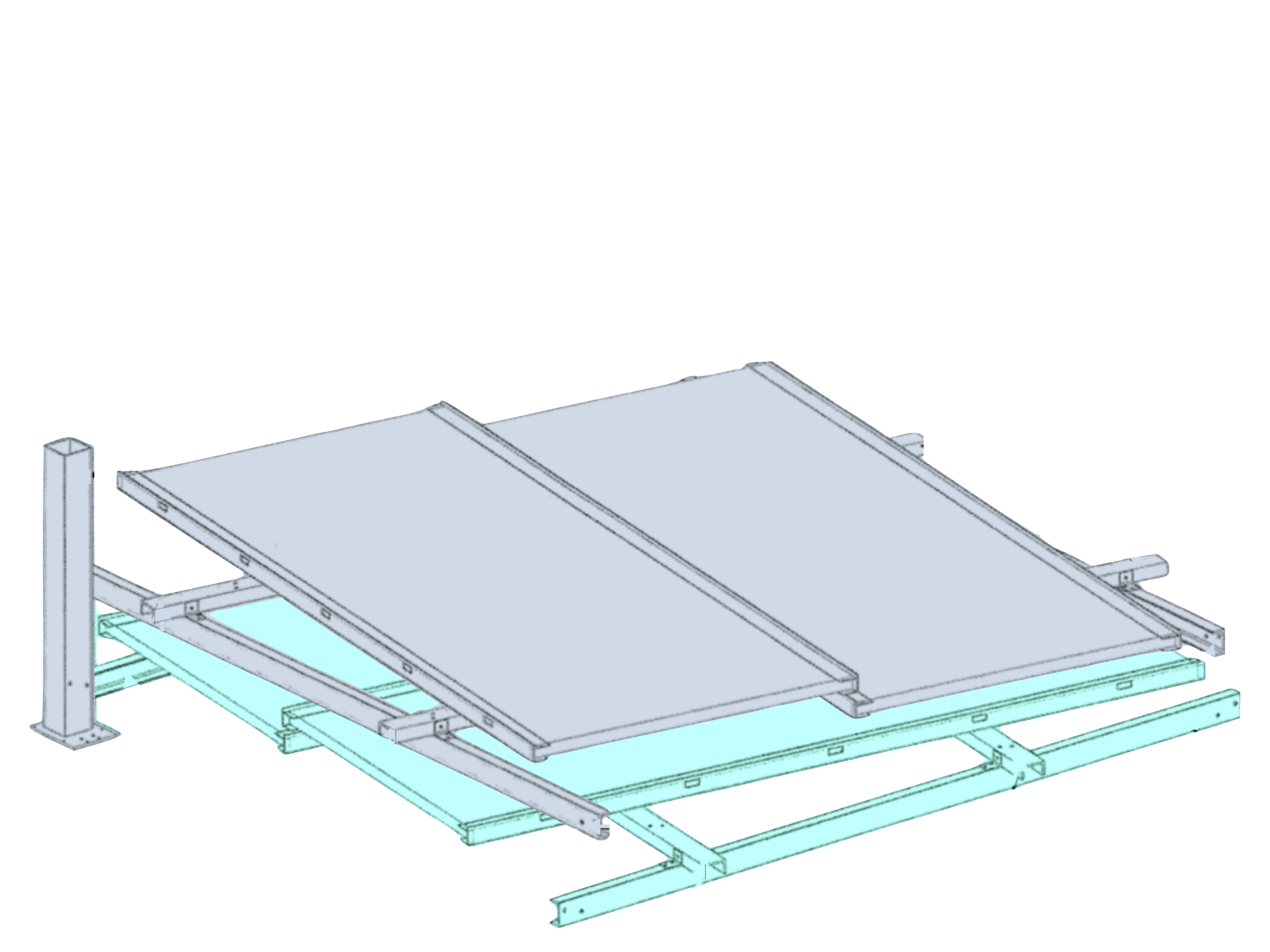}
\end{center}
\caption{
Layout of one of the RPC double layers illustrating the support structure that ensures overlap of individual chambers to avoid gaps in coverage.
}
\label{Fig:RPCassembly}
\end{figure}

The RPCs were arranged into six layers, each consisting of two chambers placed side by side. 
The RPCs in a layer were vertically offset by 10 cm to allow for overlap in order to avoid gaps in coverage.
The layers were grouped into three double-layers, which were each composed of two layers horizontally rotated by $90^\circ$ relative to each other. 
The rotated strips in each double-layer provided a measurement of two orthogonal spatial coordinates in a horizontal plane.
The double-layers were also rotated slightly relative to each other.
An example of one of the double-layers is shown in Figure \ref{Fig:RPCassembly}. 

%%%%%%%%%%%%%%%%%%%%%%%%%%%%
\subsection{Electronics, trigger, and data acquisition}

The analog signal of each scintillator was split into two paths. One was sent to an analog-to-digital converter (ADC, model LeCroy 1182) to measure the charge of each pulse. The other was sent into a discriminator (LeCroy 623B), providing a logic signal. This logic signal was input to a multi-hit time-to-digital converter (TDC, model CAEN V767) to measure signal arrival times.
RPC data were acquired on receiver cards housed in a Local Station \cite{Assiro:2004jq}. 
Each card recorded the address of the strips that were hit as well as the time from the corresponding pads.

The trigger for the \mbox{MATHUSLA} test stand was based on the top and bottom scintillator layer signals.
The top (bottom) layer signal is defined as the logical OR of all the scintillation counters in the top (bottom) scintillator plane. 
There were two primary triggers, corresponding to \emph{upward}-going particles and \emph{downward}-going particles. 

The logical AND of the top and bottom layer signals with the relative timing delay expected for upward-going (downward-going) particles traveling near the speed of light provided the upward (downward) trigger. 
Additional triggers considering only single-layer information were used for crosschecks and scintillator efficiency estimates.

When a trigger was received, the digitized charge and timing information for the scintillators was stored in the buffers of the ADCs and TDC, and the Local Station transferred the RPC data to an ARGO Memory Board.
These modules were read out after each event by a PC via a VME controller board (CAEN V2718).
The PC saved the complete raw data to files on disk for offline analysis.

%%%%%%%%%%%%%%%%%

\subsection{Coordinate system}

The origin of the test stand coordinate system is defined to be the center of the entire detector, at the midpoint of the full height, width, and length of the overall support structure. The $x$-axis is parallel to the ground and is aligned with the counterclockwise direction of the LHC (approximately east). The $y$-axis is also parallel to the ground and points away from the center of the LHC (approximately south). The horizontal edges of the support structure are aligned with these axes. The $z$-axis is directly downward. In this coordinate system, the ATLAS IP is at $(x, y, z) = (2.4~\mathrm{m}, 0.0~\mathrm{m}, 83.0~\mathrm{m})$.

The strips of each RPC layer in the test stand were approximately aligned with either the $x$- or $y$-axis. Each RPC double-layer consisted of one $x$-measuring layer and one $y$-measuring layer. Based on the RPC strip width (Section \ref{sec:rpc}), the resolution of each spatial coordinate measurement is approximately 2~cm.

Tracks in the test stand are spatially parameterized by two angles. \emph{Zenith angle} refers to the smallest angle between the track and the $z$-axis (either the $+z$ or $-z$ direction) and ranges from 0$^\circ$ to 90$^\circ$. The \emph{azimuthal angle} is the 2D polar angle of the projection of the $+z$ direction of the track onto the $(x, y)$-plane and ranges from -180$^\circ$ to 180$^\circ$. The angles as defined here rely only on spatial information and do not depend on whether a track is upward-going or downward-going.

%%%%%%%%%%%%%%%%%%%%%%%%%%%%
\section{Data analysis}

\subsection{Timing calibration}

The use of timing information is crucial in reliably reconstructing good tracks and distinguishing upward-going tracks from downward-going tracks. As such, it is imperative to ensure that the timing is consistent across all detector elements in the test stand by applying appropriate timing calibrations. 
Two types of timing calibration are applied: one that addresses characteristic delays between detector elements and one that addresses time slewing. 
After applying these calibrations, the timing resolution of the scintillation counters is better than 3~ns and a typical RPC pad has a resolution better than 4~ns~\cite{RPC_timeResolution}.

\subsubsection{Characteristic delays}

There is a delay from the time when a particle hits a detector to the time when the hit is recorded.
This delay depends on characteristics such as high voltage settings, drift time within the detector, and cable lengths.
Hence, each of the 59 scintillators and 120 RPC pads is characterised by its own typical delay.
To use timing information appropriately, these delays are calibrated. 

The calibration is performed with downward-going cosmic rays.
In each event and for each possible pair of detectors that were hit, the time difference is recorded, minus the expected time of flight between them for a particle traveling at the speed of light. 
If all delays were the same for all detectors, the mean of this distribution would be centered at zero. 
In reality, the mean is the difference between the characteristic offsets of the detectors considered. 
A Gaussian is fit to the distribution of the timing difference for each pair of detectors.
The offsets are determined by performing a least-squares fit of the means of all the Gaussians. 
After all timing calibrations have been applied, a similar fit is performed for the widths of all the Gaussians in order to calculate the timing uncertainty for each scintillator and RPC pad.

\subsubsection{Time slewing correction for scintillators}
 
The time of a hit in a scintillator is determined by the instant when the voltage of the signal pulse passes a discriminator threshold.
This introduces a time slewing effect: larger pulses cross the discriminator threshold earlier than smaller pulses.
The following calibration procedure ensures consistent timing information regardless of the pulse size.

For each scintillator, the time difference between hits in the given scintillator and all other detectors (corrected for time of flight) is plotted against the integrated charge of the corresponding pulse in the scintillator.
A power function is fitted to the distribution and the parameters of that function are used to provide a time correction based on pulse charge.

\subsection{Track reconstruction}
\label{sec:tracking}

Tracks are reconstructed by a least-squares fit using spatial and timing information from the RPC and scintillator hits.
The algorithm starts by fitting all RPC and scintillator hits in an event to a straight line consistent with the speed of light. 
Tracks are identified as either upward-going or downward-going by the direction which results in the smallest $\chi^2$ value.
In order to remove detector noise and to separate hits produced by multiple particles, an iterative process is run in which the hit with the largest $\chi^2$ contribution is removed and the track is refitted to the remaining hits.
This process proceeds until all hit residuals are smaller than a given threshold, resulting in the final reconstructed track.
The entire process is then repeated with all the discarded hits to form additional tracks, and this continues until no more tracks can be formed.
Good tracks for analysis are required to contain at least one hit in the top scintillator layer, at least one hit in the bottom scintillator layer, and hits in at least four different RPC layers. 
Only upward-going (downward-going) tracks in events passing the upward (downward) trigger and failing the downward (upward) trigger are considered in the data analysis.

In general, events recorded by the test stand are very clean:  $97\%$ of the good reconstructed tracks are in events where a maximum of 2 RPC hits and a maximum of 2 scintillator hits were discarded by the tracking algorithm. 
More than $70\%$ of the good tracks are in completely clean events where no hit was discarded. 
Only 0.1\% of events with at least one good track contain more than one good track.

\begin{figure}[h]
\begin{center}
\includegraphics[width=0.41\textwidth]{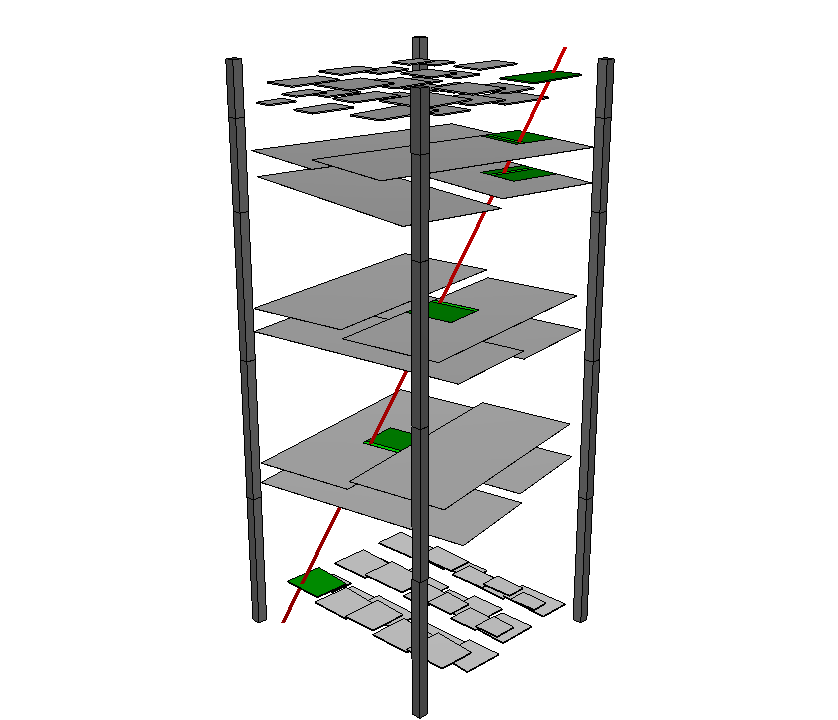}
\includegraphics[width=0.41\textwidth]{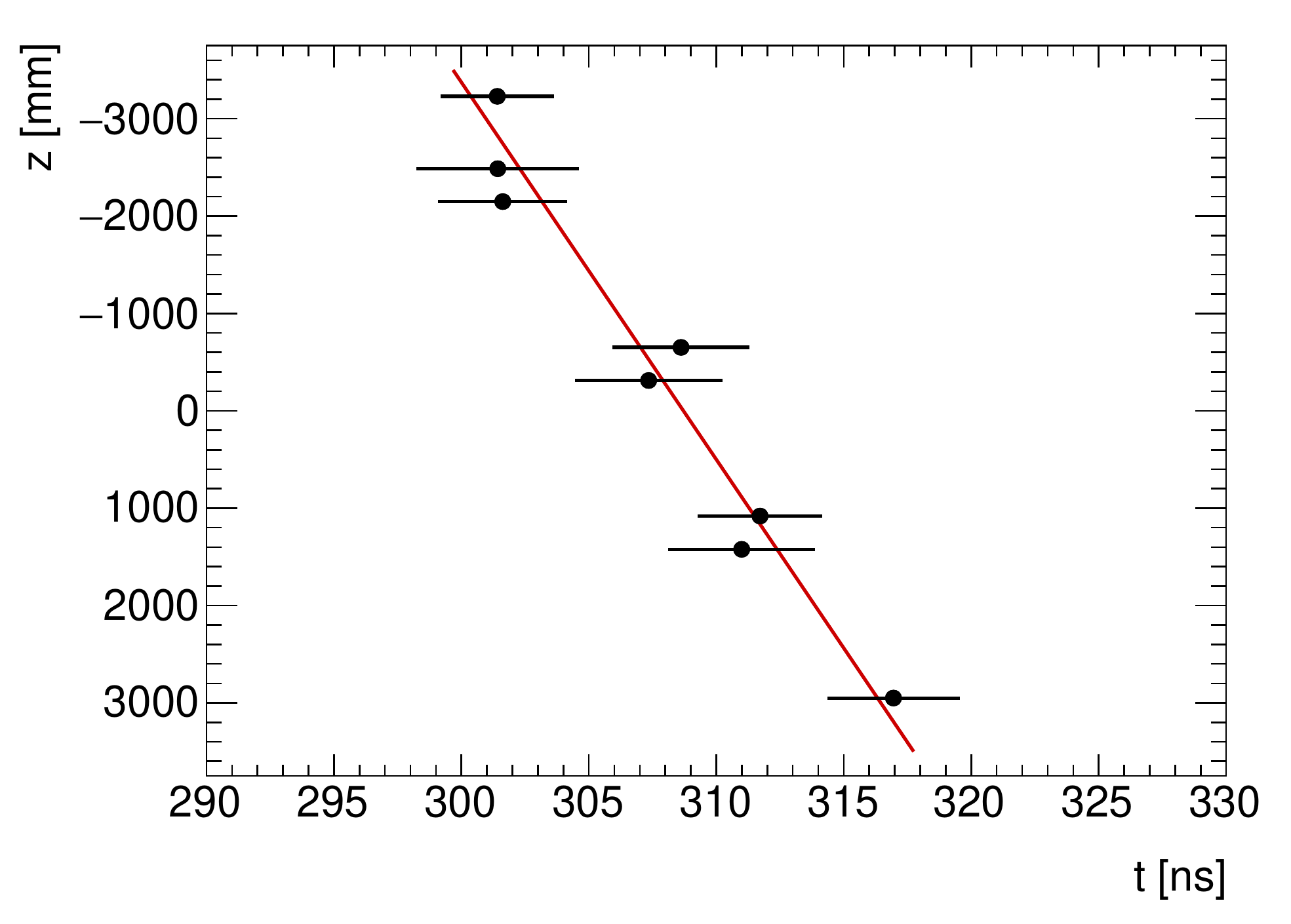}\\
\includegraphics[width=0.41\textwidth]{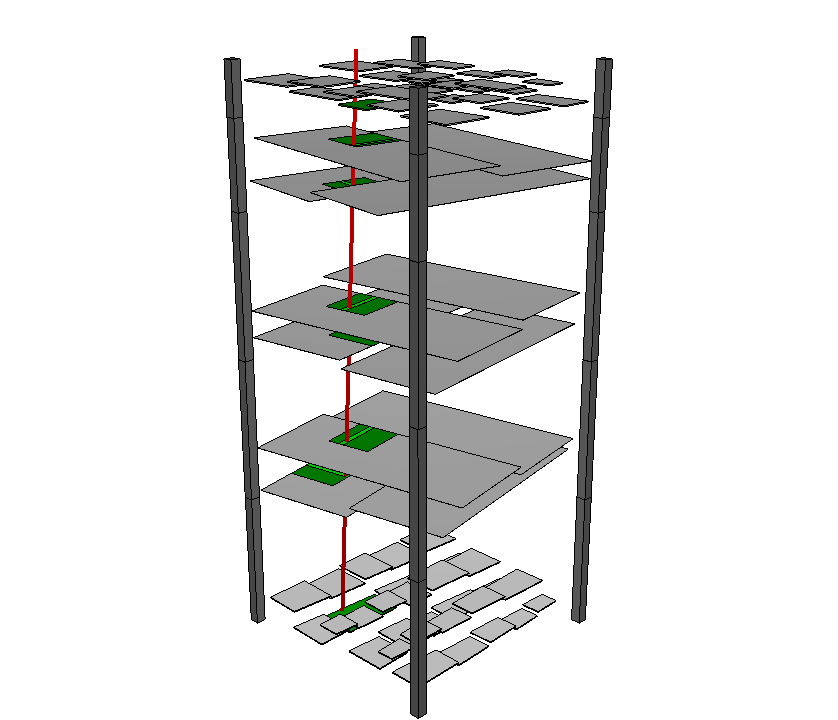}
\includegraphics[width=0.41\textwidth]{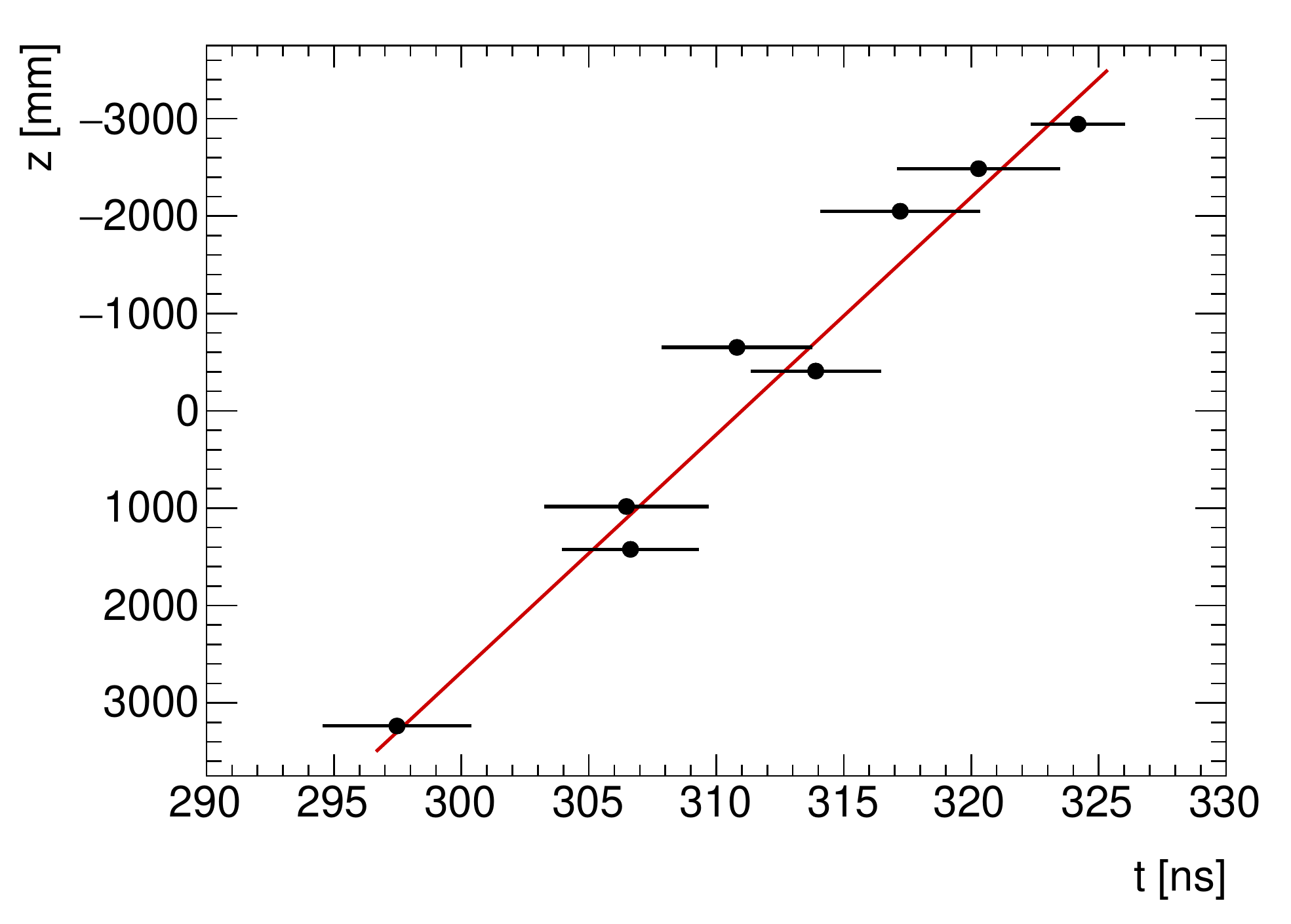}
\end{center}
\caption{
Distribution of hits for examples of a downward-going track (top) and an upward-going track (bottom) from data. Left: Event display of the scintillation counters and RPC pads comprising the track hits. Right: Plot of the $z$-coordinate versus time for each hit. The red line represents the fitted track in each case.
}
\label{f.evDisplays}
\end{figure}

Figure \ref{f.evDisplays} shows an example of a downward-going track (top) and an upward-going track (bottom) from test stand data. The left panels of the figure are event displays of the example tracks, showing the scintillation counters and RPC pads corresponding to track hits in green. The red line represents the fitted track. The right panels of the figure show the time and $z$-coordinates of the hits forming the track.

\subsection{Detector efficiency}
\label{sec:efficiency}

The efficiencies of all scintillation counters and RPC strips were estimated using the data collected by the test stand. 
RPC efficiencies were calculated using data from the primary triggers, using information from both the top and the bottom scintillator layers. 
To avoid a trigger bias in the calculation of the scintillators' efficiency, data selected by the single-layer triggers were used.
For this procedure, tracks were reconstructed as in Section~\ref{sec:tracking} but with less restrictive hit requirements to allow for potentially missing hits. If a track intercepts a given scintillator or RPC, this is considered to be an expected hit for the corresponding detector. If a hit was indeed recorded in event data from the intercepted detector, this is additionally considered to be a good hit. The efficiency of each detector is the ratio of the number of good hits to the number of expected hits. Purely geometric effects induced by this procedure on the calculated efficiencies were corrected for by performing the same procedure on simulated events.

All detectors used in the test stand are second hand or spares loaned from previous experiments. Hence, they have a large range of efficiencies depending on each chamber’s history, going from dead modules to 98\% efficiency detectors. The individual scintillators have efficiencies ranging from 60\% to 98\%. RPCs have efficiencies ranging from 55\% to 85\% except for two of them, containing dead modules, with lower efficiency. It is important to highlight that all simulation work reported in this paper uses the measured efficiencies of each RPC and scintillator.

%%%%%%%%%%%%%%%%%

\section{Simulation of events in the test stand}

\begin{figure}[h]
\begin{center}
\includegraphics[width=0.6\textwidth]{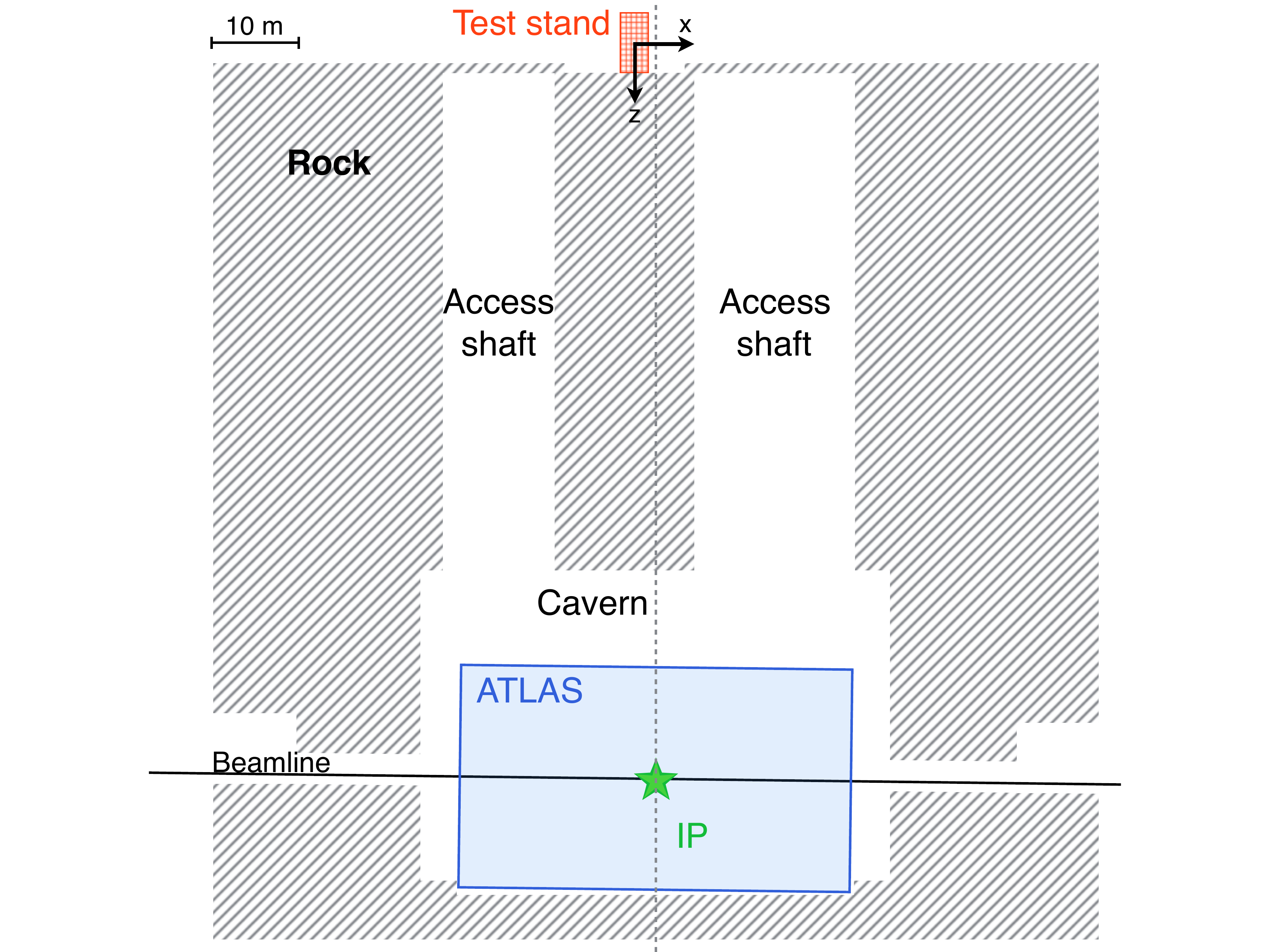}

\end{center}
\caption{
Diagram with the geometry and material of the ATLAS cavern, the test stand, and its surroundings. 
}
\label{fig:geometry}
\end{figure}

The geometry and material of the ATLAS cavern and the test stand and its surroundings as shown in Figure~\ref{fig:geometry} (to scale) were modeled with \textsc{Geant4} 10.6 \cite{Agostinelli:2002hh}. 
Starting from the ATLAS IP (green star in the diagram), the material of the ATLAS detector (blue box in the diagram), equivalent to approximately 11 nuclear interaction lengths, is simulated by introducing a 1.85 m-thick cylindrical shell of iron.
The rock (gray hatched area in the diagram) surrounding the ATLAS cavern was approximated by 45.30~m of sandstone, 18.25~m of marl, and 36.45~m of an equal mixture of sandstone and marl, as determined from a geological survey \cite{GeoSurvey1,GeoSurvey2}.
Both the scintillation counters and the RPCs, as well as parts of the supporting structure, are included in the simulation.  
Air, comprised primarily of a standard admixture of nitrogen and oxygen, fills the gaps.
The position of the test stand (red crosshatched area in the diagram) relative to the IP was determined by a combination of direct measurements of the test stand inside the SX1 building and engineering drawings of SX1 and the ATLAS cavern.

The energy deposited by an ionizing particle passing through a detector element in \textsc{Geant4} is saved as several small deposits within a few nanoseconds. These energy deposits in scintillators and RPC gas were integrated to form candidate detector hits.
Each candidate hit is assigned an efficiency, corresponding to the one calculated in Section \ref{sec:efficiency} for the scintillator or RPC that was hit.

\subsection{Cosmic rays}
\label{sec:cosmicsMC}

Cosmic rays are the dominant source of energetic charged particles at Earth's surface and represent the vast majority of tracks in the test stand. 
Since these particles originally arise from primary cosmic rays entering the atmosphere, they are initially directed downwards toward the surface. 
Downward cosmic rays can inelastically scatter or decay in test stand material or in the concrete floor and generate additional charged particles. 
These processes can produce particles traveling in any direction, including upwards.
This upward contribution, referred to as \emph{cosmic ray inelastic backscattering}, can activate the upward trigger of the test stand and produce upward-going tracks.

Downward cosmic ray particles, including muons, electrons, positrons, protons, neutrons, and photons, were generated by sampling energies and zenith angles from distributions predicted by PARMA4.0~\cite{PARMA3, PARMA4, PARMAsite}, an analytical model for estimating cosmic ray fluxes on Earth. 
The particles were simulated in the \textsc{Geant4} model of the test stand and detector hits were recorded.
In order to study both downward tracks from incoming cosmic ray particles and upward tracks produced by secondary particles with reasonable statistics, two separate sets of simulations were run. 
For the first set, initial cosmic ray particles were uniformly spatially distributed in \textsc{Geant4} just above the top scintillator plane of the test stand within a horizontal square area of 4.6~$\times$~4.6~m$^2$.  
To study upward tracks generated by secondary particles, the initial cosmic ray particles were uniformly distributed 0.6 m above the bottom of the test stand within a horizontal square area of 6.2~$\times$~6.2~m$^2$.
This height was chosen so as to be just above the raised concrete floor adjacent to the test stand. 
The horizontal area chosen for the initial distribution of particles in the first (second) set of simulations ensured that any point on a top (bottom) scintillator received 95\% or more of the total cosmic ray flux.

\subsection{LHC $pp$ collisions}
\label{sec:ppMC}

The dominant LHC $pp$ collision processes that can produce particles reaching the test stand are the production of $W$, $Z$, $\ccbar$, $\bbbar$, and $\ttbar$. 
Muons from the immediate or sequential decays of these particles are the main source of hits from $pp$ collisions in the test stand.

Direct simulation of the above processes at $\sqrt{s}$~=~13~TeV by \textsc{Pythia} 8.2 \cite{Sjostrand:2014zea} was used to estimate the acceptance, which is defined as the number of events in which \textsc{Geant4} records a sufficient collection of candidate hits to reconstruct a good upward track (as defined in Section \ref{sec:tracking}) divided by the total number of events generated.
From the acceptance, a raw track rate can be computed that does not account for the efficiency of the detector elements. 

The processes for $W$, $Z$, and $\ttbar$ were normalized to their measured cross-sections at 13~TeV \cite{ATLAS:2016601, PhysRevLett.116.052002}. The processes $\ccbar$ and $\bbbar$ were simulated with a minimum $p_\mathrm{T}$ threshold on the leading outgoing parton of 25~GeV. Lower $p_\mathrm{T}$ thresholds give a very small contribution to the test stand rate because the minimum $p_\mathrm{T}$ a muon must have in order to reach the surface is approximately 30~GeV.
Both ATLAS and CMS have measurements of the inclusive $b$-jet cross-section at 7~TeV \cite{Aad2011:bjet, Chatrchyan2012:bjet}, but measurements at 13~TeV were not available.
Although neither experiment measures the cross-section in a region of phase space directly relevant to the test stand, in the most relevant regions of phase space \textsc{Pythia} over-predicts the data by approximately 25\%. Thus a 25\% systematic uncertainty was assigned to the cross-section for $\bbbar$ and $\ccbar$.

There are considerable uncertainties in the material composition between the test stand and the IP.  What is simulated is an approximation of a nearby geological survey, but the magnitude of local variations in the material is unknown.  Furthermore, a small change in the total material can lead to a large change in the rate of upward-going muons because the initial momentum spectrum decreases steeply with increasing momentum.  For instance, varying the depth of the rock between the IP and the test stand by $\pm1$~m (a 2\% change in the total rock) results in a 5--10\% change in the rate of muons, depending on the process.  The rate of muons from $W$ bosons in particular is sensitive to the material description because the momentum distribution for these muons peaks around 40~GeV, which is in a range for which the survival probability to reach the surface is changing quickly.  
The material uncertainty leads to a 5--10\% systematic uncertainty on the predicted rate of muons.

Normalizing the rates to an instantaneous luminosity of $10^{34}$~cm$^{-2}$~s$^{-1}$ results in an expected total raw track rate of $ 12.3 \pm 1.4 $ per hour, including systematic uncertainties, before applying detector inefficiencies.
These results are summarized in Table \ref{t.acceptance}.

\begin{table}[ht]
\centering
\caption{Summary of cross-sections, acceptances, and expected rates  at the test stand from various LHC $pp$ collision processes at $\sqrt{s}$ = 13~TeV.
The rates are normalized to an instantaneous luminosity of $10^{34}$~cm$^{-2}$~s$^{-1}$ and do not include test stand detector inefficiencies.}
\label{t.acceptance}
\vspace{0.5cm}
\begin{tabular}{cccc}
\hline
Process & Cross-section [nb] & Acceptance $\left(\times 10^{-6}\right)$ & Raw track rate [hr$^{-1}$] \\
\hline
$ W \rightarrow \mu \nu $ & \multirow{2}{*}{$ 20.6 \pm 0.7 $} & $ 6.10 \pm 0.17 $ & $ 4.5 \pm 0.4 $  \\
$ W \rightarrow \tau \nu $ & & $ 0.57 \pm 0.02 $ & $ 0.42 \pm 0.04 $ \\
\hline
$ Z \rightarrow \mu \mu $ &  \multirow{2}{*}{$ 1.98 \pm 0.06 $} & $ 17.3 \pm 0.4 $ & $ 1.23 \pm 0.11 $ \\
$ Z \rightarrow \tau \tau $ & & $ 1.59 \pm 0.04 $ & $ 0.11 \pm 0.01 $ \\
\hline
$ \ccbar $ & $ 5600 \pm 1400 $ & $ 0.0052 \pm 0.0007 $ & $ 1.1 \pm 0.3 $  \\
\hline
$ \bbbar $ & $ 5300 \pm 1300 $ & $ 0.0257 \pm 0.0016 $ & $ 4.9 \pm 1.3 $ \\
\hline
$ \ttbar $ & $ 0.75 \pm 0.09 $ & $ 4.64 \pm 0.12 $ & $ 0.12 \pm 0.02 $ \\
\hline
Total & --- & --- & $ 12.3 \pm 1.4 $ \\
\hline
\end{tabular}
\end{table}

%%%%%%%%%%%%%%%%%%%%%%%%%%%%
\section{Results}

The track rates and distributions from data are compared to expectations from simulation after applying detector efficiencies. 
Data are separated into periods when no beam was present in the LHC (runs with no beam) and periods when there were beams circulating in the LHC (runs with beam).
Only good tracks as defined in Section \ref{sec:tracking} are included in these studies.

\begin{figure}[h]
\begin{center}
\includegraphics[width=0.45\textwidth]{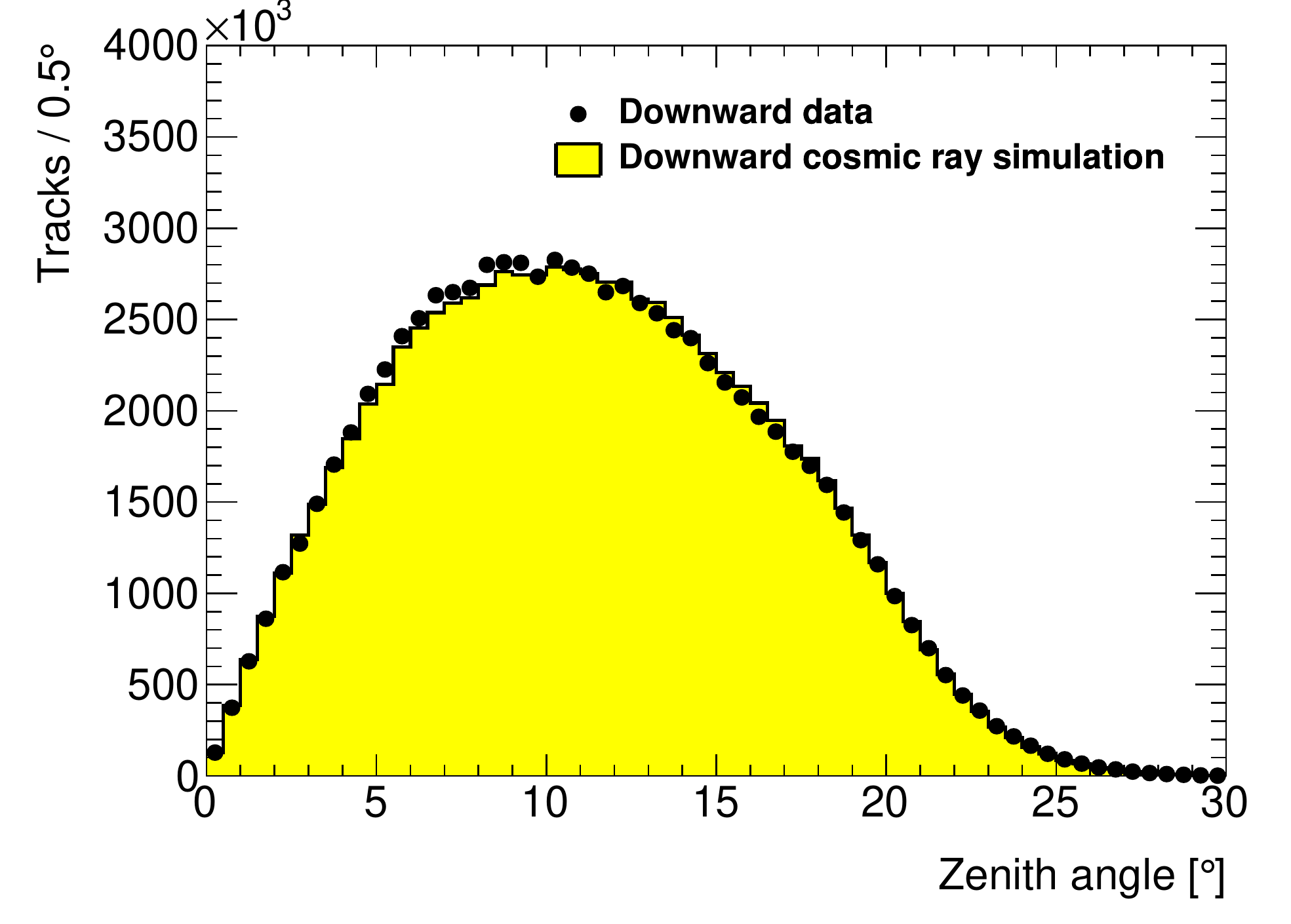}
\includegraphics[width=0.45\textwidth]{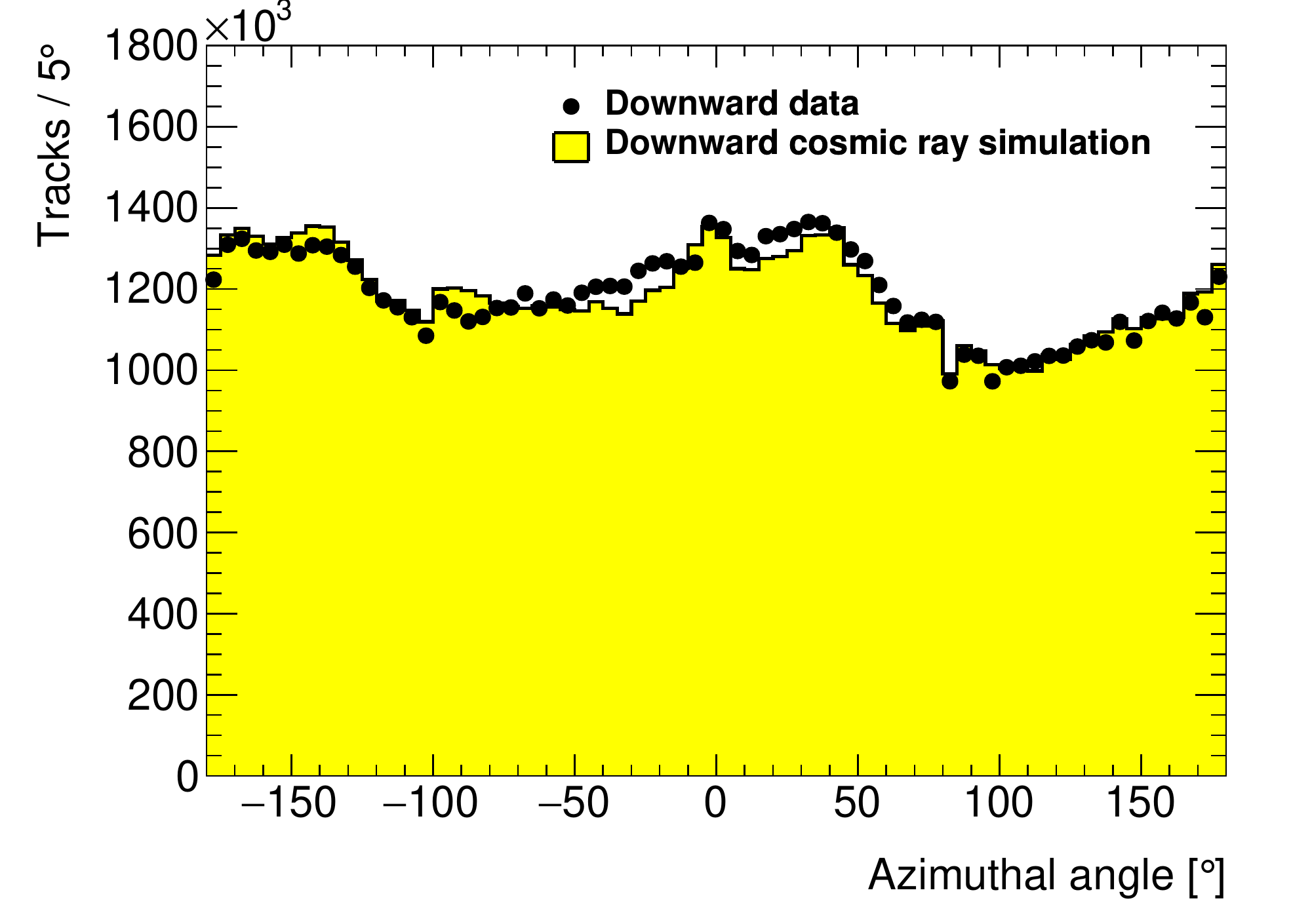}
\end{center}
\caption{
Distribution of reconstructed downward-going tracks as a function of the zenith angle (left) and the azimuthal angle (right). Data events are shown as black markers. The yellow area corresponds to the downward cosmic ray simulation, normalized to data.
}
\label{fig:down_data_MC}
\end{figure}

Figure~\ref{fig:down_data_MC} shows the zenith angle (left) and azimuthal angle (right) distributions of downward tracks in all data and the expected distributions from the downward cosmic ray simulation, normalized to data. 
The good agreement between these sets of events confirms that the downward-going tracks are properly reconstructed.

The number of downward cosmic ray tracks increases with zenith angle from 0$^\circ$ to 10$^\circ$ due to the increasing solid angle.
The distribution peaks at about 10$^\circ$ and decreases at higher zenith angles as the geometric acceptance of the test stand diminishes. The fluctuations in the number of tracks as a function of azimuthal angle also reflects the geometric acceptance of the test stand.

\begin{figure}[h]
\begin{center}
\subfigure[]{\includegraphics[width=0.45\textwidth]{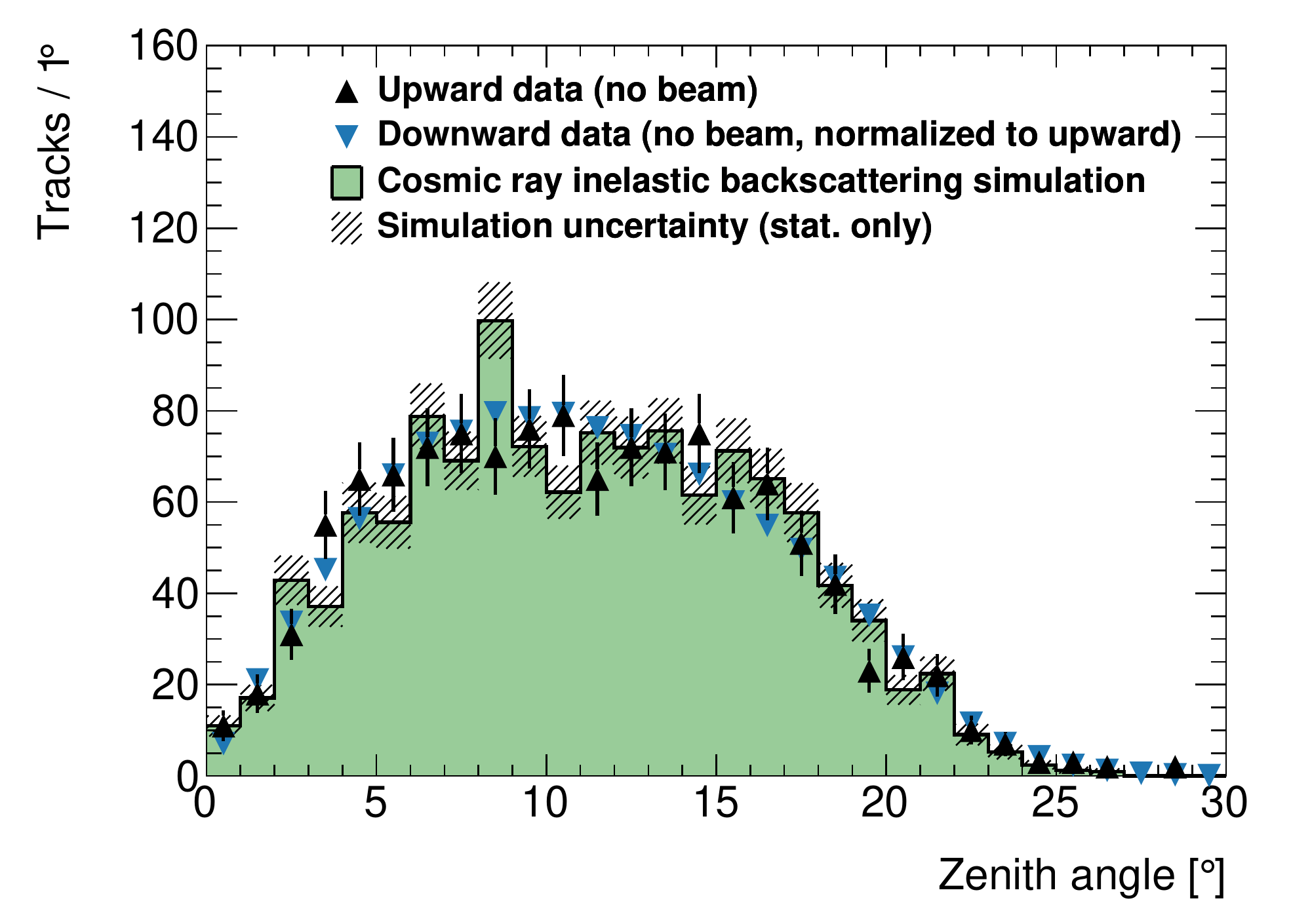}}
\subfigure[]{\includegraphics[width=0.45\textwidth]{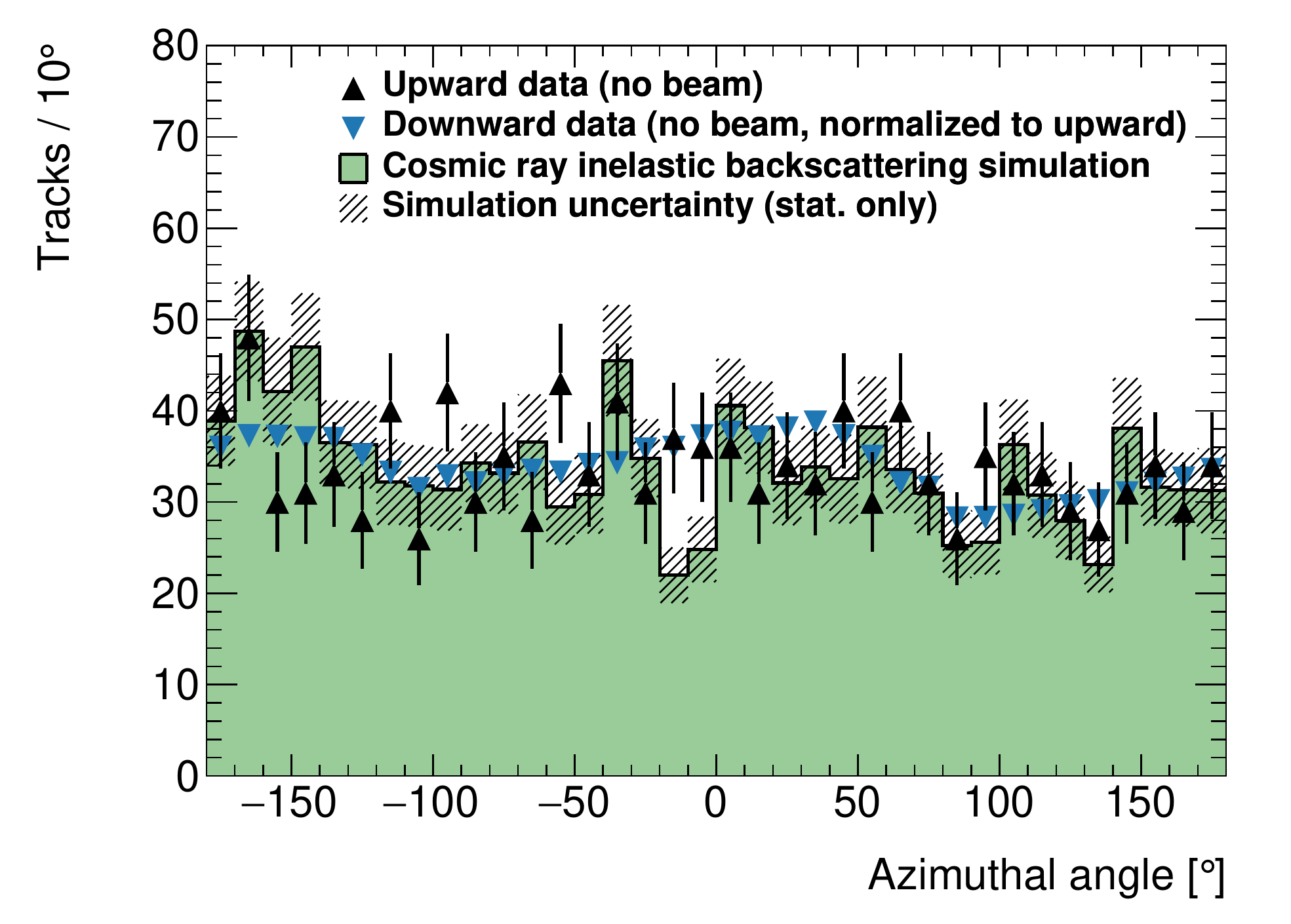}}
\end{center}
\caption{
Distribution of reconstructed upward-going tracks as a function of the zenith angle (left) and the azimuthal angle (right). Data events with no beam are shown as upward-pointing black triangles.  The green area corresponds to the cosmic ray inelastic backscattering simulation, with the statistical uncertainty shown by the hatched area. This distribution is normalized by multiplying it by the ratio of downward tracks in simulation to downward tracks in data without beam. Downward-going tracks from data events with no beam, normalized using the $R_\mathrm{up-to-down}$ factor, are included in the plot as downward-pointing blue triangles.
}
\label{fig:upTracksdata_albedoTracksMC}
\end{figure}

In the case of upward-going tracks, there are two components to take into account. The first component comes from cosmic ray inelastic backscattering. 
Upward particles generated by cosmic rays can be produced by cosmic muon decays or by interactions with material in the test stand or in the floor of the SX1 building. 
These upward particles are emitted across the entire geometric acceptance of the test stand, as seen in Figure~\ref{fig:upTracksdata_albedoTracksMC}, where upward tracks in data with no beam are shown as upward-pointing black triangles. The angular distributions for downward tracks from the same dataset, normalized to the number of upward tracks, are shown as downward-pointing blue triangles.
These upward and downward data distributions are consistent within uncertainties. The final angular distributions of these tracks are dominated by the narrow geometric acceptance of the test stand and are insensitive to the initial angular distributions of the interacting particles.

From these runs without beam, a ratio that relates the rate of upward inelastic backscattering to the rate of incident downward cosmic rays is obtained:

\begin{linenomath*}
\begin{equation*}
    R_\mathrm{up-to-down} = \frac{(\mathrm{Number~of~upward~tracks})_{\mathrm{data,~no~beam}}}{(\mathrm{Number~of~downward~tracks})_{\mathrm{data,~no~beam}}} = (7.0 \pm 0.2) \times 10^{-5},
\end{equation*}
\end{linenomath*}

where purely statistical uncertainties are shown.

Figure~\ref{fig:upTracksdata_albedoTracksMC} also compares the data distributions to the results of the cosmic ray inelastic backscattering simulation. The statistical uncertainty of the simulation is shown by the hatched area. The predicted rate and angular distributions from this simulation are compatible within uncertainties with the observed data.

The second component of upward tracks comes from muons created in LHC $pp$ collisions. Since the test stand operated almost directly 80 m above the ATLAS IP, tracks from these muons are expected to be concentrated at small zenith angles.

\begin{figure}[h]
\begin{center}
\subfigure[]{\includegraphics[width=0.45\textwidth]{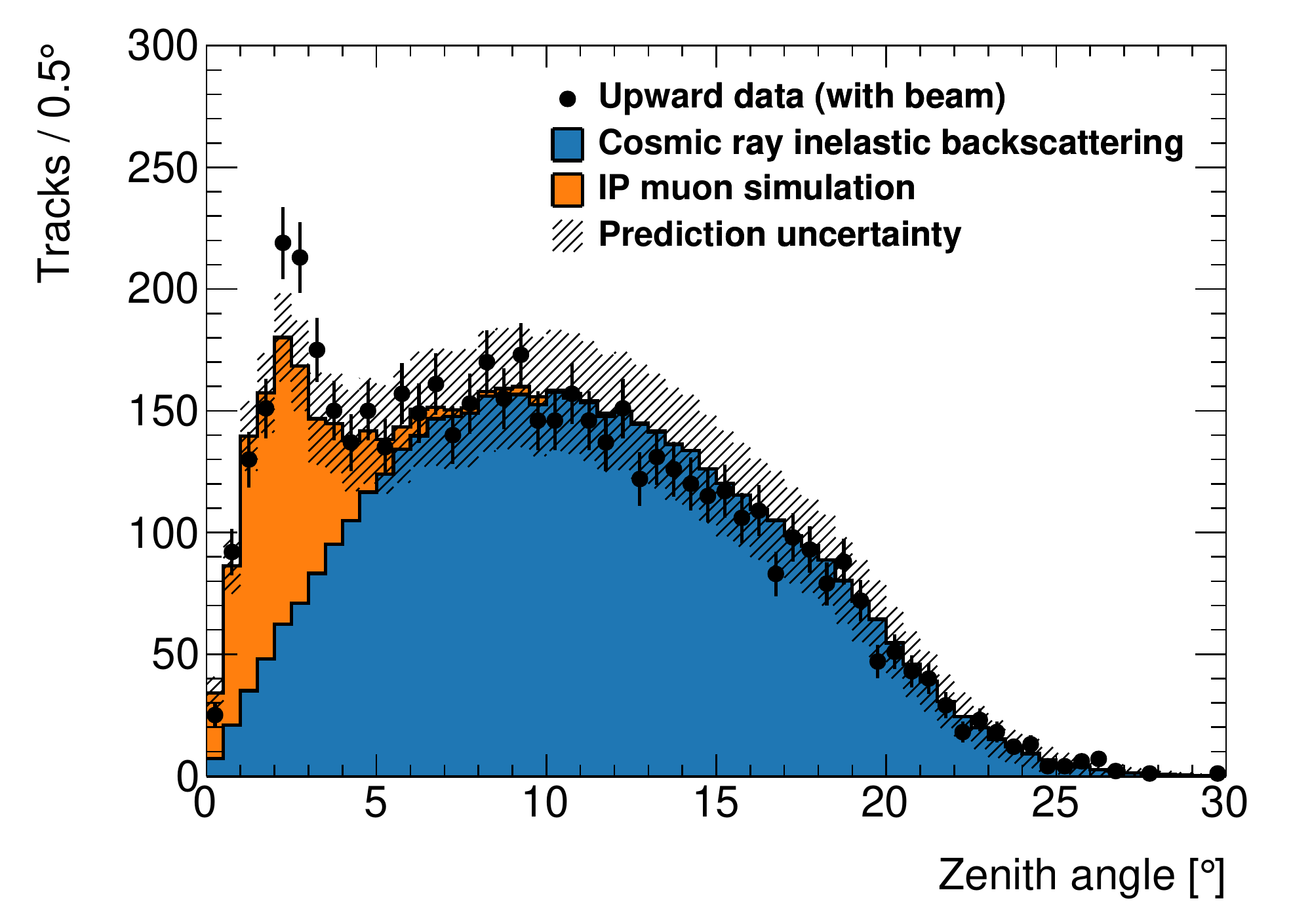}\label{fig:uptracks_beam_zenith}}
\subfigure[]{\includegraphics[width=0.45\textwidth]{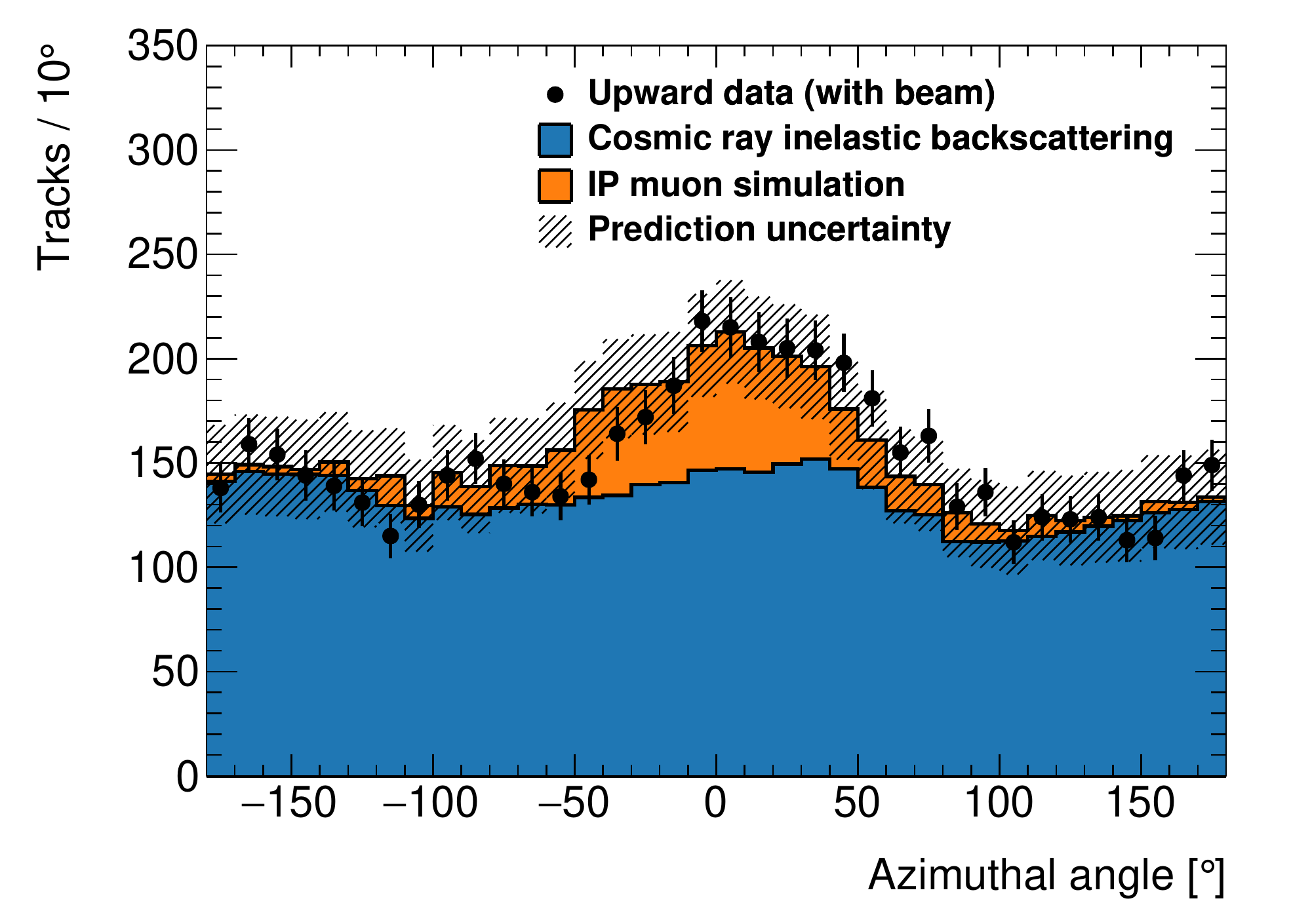}}
\end{center}
\caption{
Distribution of reconstructed upward-going tracks as a function of the zenith angle (left) and the azimuthal angle (right). Data events in runs with beam are shown as black markers.
Overlaid is a simulation of particles coming from the ATLAS IP in orange. The blue distribution corresponds to cosmic ray inelastic backscattering, showing the downward-going tracks from data runs with beam, normalized using the $R_\mathrm{up-to-down}$ factor.
The hatched area shows the combination of the uncertainties of the IP muon simulation and the cosmic ray inelastic backscattering prediction. 
}
\label{fig:uptracks_beam}
\end{figure}

Figure~\ref{fig:uptracks_beam} shows the angular distributions of upward tracks in data with beam (black markers). 
The blue area is the prediction for tracks from cosmic ray inelastic backscattering that is derived from the distribution of downward tracks reconstructed in data with beam, normalized by the ratio $R_\mathrm{up-to-down}$ defined above. Given that both angular distributions are identical, downward tracks are used in this normalization to avoid large statistical fluctuations from the smaller upward tracks dataset.
The orange area in this figure shows the expected tracks from the simulation of muons produced in LHC collisions.
The predicted track distributions from this simulation, after accounting for detector efficiencies, are normalized by the total integrated luminosity reported by ATLAS during the test stand runs with beam.
The hatched area in this plot shows the combination of uncertainties of the cosmic ray inelastic backscattering prediction and the IP muon simulation.

\begin{figure}[h]
\begin{center}
\subfigure[]{\label{fig:downTracks_rates}\includegraphics[width=0.45\textwidth]{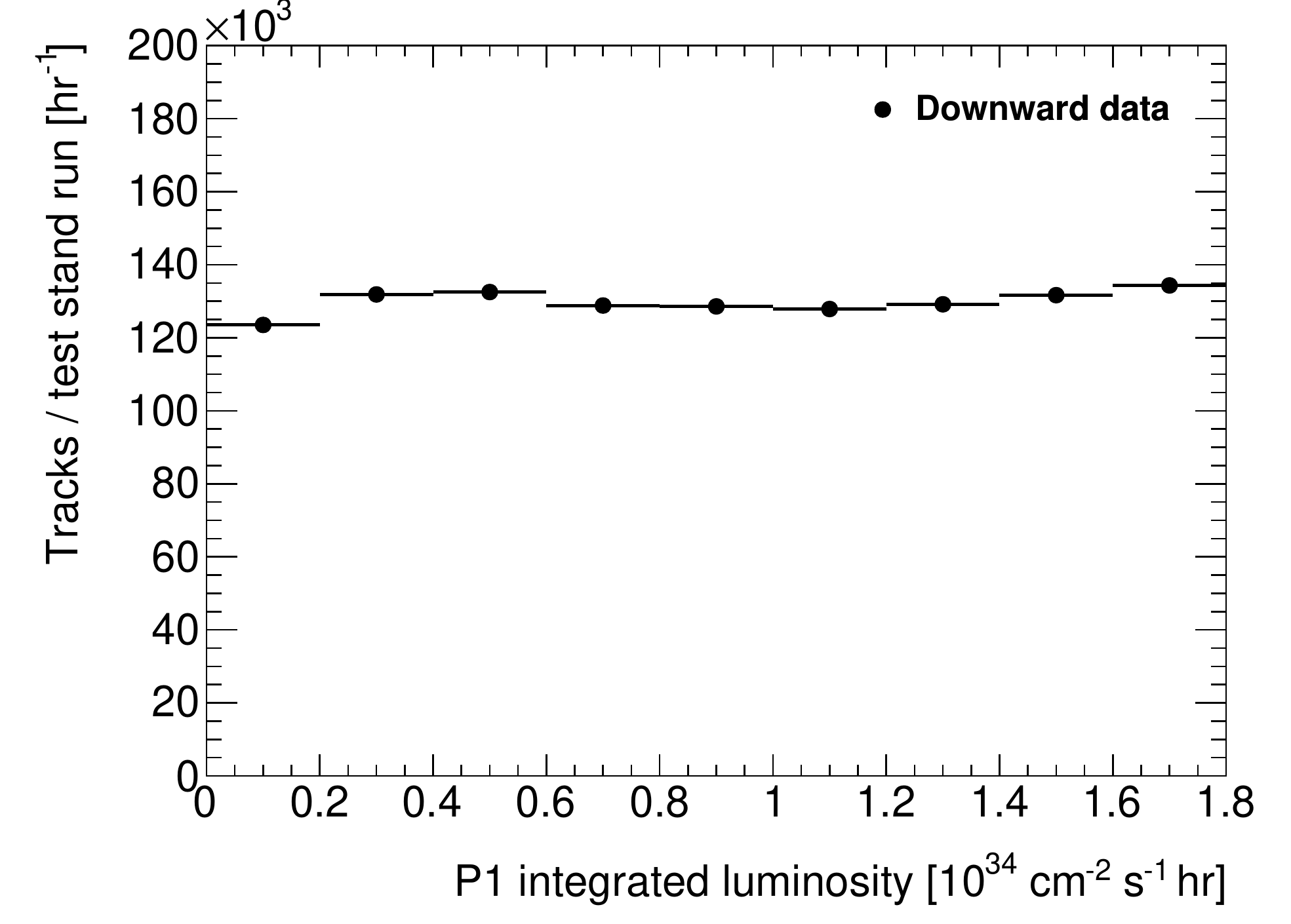}}
\subfigure[]{\label{fig:upTracks_rates}\includegraphics[width=0.45\textwidth]{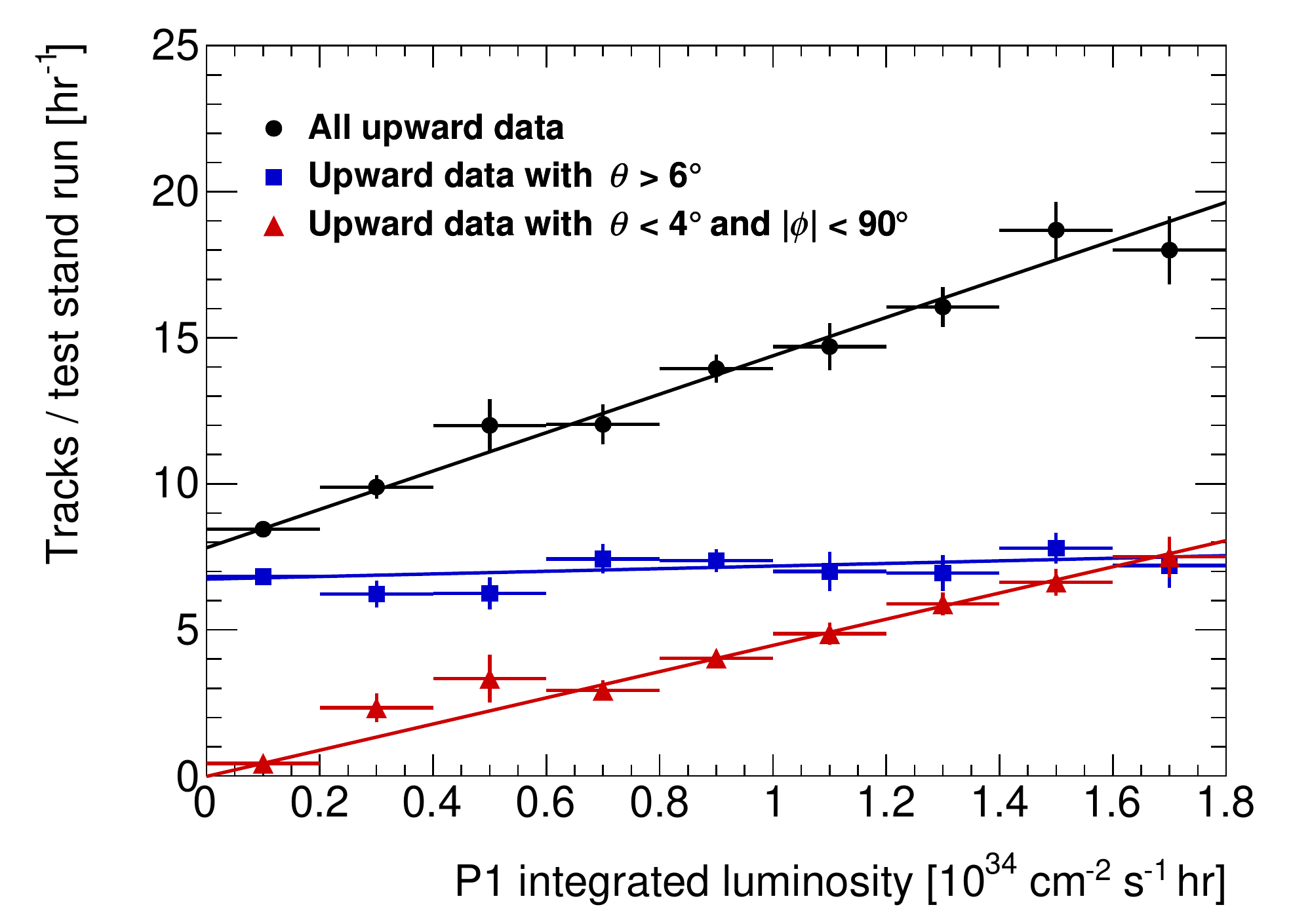}}
\end{center}
\caption{
Distribution of the number of reconstructed tracks as a function of the ATLAS integrated luminosity during each one-hour test stand run. Left: Downward tracks. Right: Upward tracks (black circles), including tracks with a zenith angle ($\theta$) $ > 6^\circ $ (blue squares) and tracks with a zenith angle $ < 4^\circ $ and absolute value of azimuthal angle ($\phi$) $ < 90^\circ $ (red triangles).
}
\end{figure}

The number of tracks per hour was studied as a function of luminosity. Figure~\ref{fig:downTracks_rates} shows that the rate of downward tracks is independent of the luminosity, as expected from cosmic rays.
The small fluctuations in the number of tracks for different luminosity points are caused by fluctuations in RPC efficiency.
Figure~\ref{fig:upTracks_rates} shows the rates for upward tracks. 
Black circles show the rate for all upward tracks. 
Blue squares represent the tracks with a zenith angle ($\theta$) greater than $6^\circ$ and correspond to the majority of the inelastic backscattering from cosmic rays. As expected, this distribution is independent of the luminosity at the LHC.
Tracks with a zenith angle less than $4^\circ$ and an absolute value of azimuthal angle ($\phi$) less than $90^\circ$ correspond mainly to particles coming from LHC $pp$ collisions. These are shown by the red triangles, where a clear trend is observed with the rate increasing linearly as the luminosity increases. 
This confirms that the peak of upward tracks at small zenith angles is correlated with beam activity. 
A linear fit to these points was performed, providing the following result, in which the quoted uncertainties are purely statistical.

\begin{linenomath*}
\begin{equation*}
\textrm{Upward tracks (} \theta < 4^\circ \textrm{, } |\phi| < 90^\circ \textrm{)} = (4.48 \pm 0.16) \times \left( \frac{\textrm{Integrated luminosity}}{10^{34}~\mathrm{cm}^{-2}~\mathrm{s}^{-1}~\mathrm{hr}} \right) + (-0.02 \pm 0.03)
\end{equation*}
\end{linenomath*}

The positive slope and intercept near zero are strong evidence that tracks in this solid angle selection are predominantly coming from LHC particles.

In order to compare the rate of upward-going tracks between simulation and the data, tracks with a zenith angle less than 10$^\circ$ are selected. This selection is different than what is used in the fit estimate and is deliberately loose in order to avoid uncertainties from potential mismodeling of the zenith angle distributions that might bias the result.  After accounting for detector inefficiency, the predicted rate of IP muon tracks from all sources with a reconstructed zenith angle less than $10^\circ$ is $4.8\pm0.5$ per $(10^{34}~\mathrm{cm}^{-2}~\mathrm{s}^{-1}~\mathrm{hr})$.  A  rate of $5.7\pm0.7$ IP muon tracks per $(10^{34}~\mathrm{cm}^{-2}~\mathrm{s}^{-1}~\mathrm{hr})$ is measured after subtracting the expected number of tracks from cosmic ray inelastic backscattering.  This is compatible with the predicted rate within the known uncertainties.

\section{Conclusions}

The data recorded by the MATHUSLA test stand in 2018 during periods both with and without LHC collisions are dominated, as expected, by downward-going cosmic rays. Upward-going tracks, identified by timing, have two components. 
One is background from cosmic ray inelastic backscattering that has an observed angular distribution consistent with the observed downward cosmic ray angular distribution because both are determined by detector acceptance.
The second source of observed upward-going tracks is shown to be consistent with expected muons from LHC collisions, which have a significantly narrower angular distribution that is determined by the small solid angle subtended by the test stand.
The measured rate of muons from the IP scales linearly with luminosity and is consistent with Monte Carlo simulated rates. 
The test stand results confirm the background assumptions in the \mbox{MATHUSLA} proposal and demonstrate that there are no unexpected sources of background. These results give confidence in the \mbox{MATHUSLA} projected physics reach. 

\section{Acknowledgments}

We thank CERN for the successful operation of the LHC.
We thank the ATLAS collaboration for allowing us to install our apparatus in the ATLAS assembly hall and Karl Jacob and Ludovico Pontecorvo for their encouragement and support. We are grateful to the CMS collaboration for their substantial technical support and key assistance during the installation of the test stand and to Dmitri Denisov for the loan of the D\O\, scintillation counters. We thank Luigi Di Stante for his crucial contribution to the assembly of the RPC layers in the test stand and Kacper Kapusniak for installing and commissioning the test stand RPC gas system. We are grateful to Juan Carlos Ortega for helpful suggestions. YS and EE thank the PAZY foundation for their financial support. HL and GW thank the University of Washington Royalty Research Foundation for their support. JPC and SAT thank the Rutgers School of Arts \& Sciences Dean's office for partial support of this work.The simulations and analysis in this work were facilitated through the use of advanced computational, storage, and networking infrastructure provided by the Hyak supercomputer system at the University of Washington.

%%%%%%%%%%%%%%%%%
\bibliography{references}
\bibliographystyle{JHEP}
%%%%%%%%%%%%%%%%%

\end{document}